\documentclass[a4paper,12pt]{article}
\pdfoutput=1
\usepackage{jcappub}
\usepackage{amsthm,graphicx}
\usepackage{epsfig}
\usepackage{latexsym, amssymb} 
\usepackage{amsmath}
\usepackage[normalem]{ulem}
\usepackage{xcolor}
\usepackage{mathtools}
\usepackage{accents}
\usepackage{rotating}
\hypersetup{
    colorlinks=true,
    linkcolor=blue,
    filecolor=magenta,      
    urlcolor=blue,
    breaklinks=true
 }
\def\beq{\begin{equation}}
\def\eeq{\end{equation}}
\def\br{\begin{eqnarray}}
\def\er{\end{eqnarray}}
\def\benu{\begin{enumerate}}
\def\efnu{\end{enumerate}}
\def\nn{\nonumber}

\def\l{\left}
\def\r{\right}

\def\cl{{\cal C}_{\ell}}

\begin{document}

\title{Exploring the discrepancy between Planck PR3 and ACT DR4}

\author[1,2,3]{Dhiraj Kumar Hazra}
 \author[4]{Benjamin Beringue}%
 \author[4]{Josquin Errard}%
\author[5,6]{Arman Shafieloo}
\author[7,8,9,10]{George F. Smoot}

\affiliation[1]{The Institute of Mathematical Sciences, CIT Campus, Chennai 600113, India}
 \affiliation[2]{Homi Bhabha National Institute, Training School Complex, Anushakti Nagar, Mumbai 400094, India}
 \affiliation[3]{INAF/OAS Bologna, Osservatorio di Astrofisica e Scienza dello Spazio, Area della ricerca CNR-INAF, via Gobetti 101, I-40129 Bologna, Italy}
 \affiliation[4]{Université Paris Cit\'e, CNRS, Astroparticule et Cosmologie, F-75013 Paris, France}
\affiliation[5]{Korea Astronomy and Space Science Institute, Daejeon 34055, Korea}%
\affiliation[6]{University of Science and Technology, Daejeon 34113, Korea}%
\affiliation[7]{Donostia International Physics Center DIPC, Basque Country, San Sebastian, Spain}
\affiliation[8]{Institute for Advanced Study, Hong Kong University of Science and Technology, Clear Water Bay, Kowloon 999077, Hong Kong, {\it emeritus}}
\affiliation[9]{Universit\'e Sorbonne Paris Cit\'e, Laboratoire APC-PCCP, Universit\'e Paris Diderot, 10 rue Alice Domon et Leonie Duquet, CEDEX 13, 75205 Paris, France, {\it emeritus}}
\affiliation[10]{Department of Physics and LBNL, University of California, MS Bldg 50-5505 LBNL, 1 Cyclotron Road, Berkeley, CA 94720, USA, {\it emeritus}}
\emailAdd{dhiraj@imsc.res.in, beringue@apc.in2p3.fr, josquin@apc.in2p3.fr, shafieloo@kasi.re.kr, gfsmoot@lbl.gov }
\abstract
{We explore the scales and the extent of disagreement between {\it Planck} PR3 and Atacama Cosmology Telescope (ACT) DR4 data. {\it Planck} and ACT data have substantial overlap in the temperature anisotropy data between scales corresponding to multipoles $\ell\simeq 600-2500$ with complementing coverage of larger angular scales by {\it Planck} and smaller angular scales by ACT. Since the same cosmology should govern the anisotropy spectrum at all scales, we probe this disagreement in the primordial power spectrum. We use a parametric form of power law primordial spectrum that allows changes in the spectral tilt. We also reconstruct the primordial spectrum with a non-parametric method from both {\it Planck} and ACT temperature data. We find the disagreement exists within scales 0.08 -- 0.16 ${\rm Mpc}^{-1}$ where ACT temperature data prefers a scale invariant/blue spectrum. At scales larger and smaller than this window, ACT data strongly prefers a red tilt, which is consistent with {\it Planck}. This change in the spectral tilt can be identified in the ACT data at 2$\sigma$ C.L. without using {\it Planck} data, indicating that the tension is driven by different preferences for tilts within the ACT data. The addition of {\it Planck} data up to intermediate scales ($\ell\le650$) increases this significance to 3$\sigma$. Given the large overlap between {\it Planck} and ACT within 0.08 -- 0.16 ${\rm Mpc}^{-1}$ and considering the internal consistency between different {\it Planck} temperature and polarization spectra, the scope of new physics as a solution to the tension remains limited. Our results --- a strong preference for an intermediate transition in spectral tilt and the variation of this preference in different data combinations --- indicate that systematic effects can be misperceived as new physics emerging from different non-standard cosmological processes.}

\maketitle

\section{Introduction}
Tension between different datasets are often ascribed to new physics. While in certain cases, new physics may emerge from these tensions, in many cases unknown systematic effects turn out to be the reason for the tensions. After the 4th release of ACTPOL CMB data in~\cite{ACT:2020}, a discrepancy between {\it Planck}~\cite{Planck:2018param} and ACT in estimating the spectral tilt was reported at 3$\sigma$ level~\footnote{The overall discrepancy was reported to be at the level of 2.6$\sigma$ in~\cite{Handley:2020hdp} (see also,~\cite{DiValentino:2022rdg})}. Since ACT does not detect the first CMB peak, a prior on  the angular power spectrum at multipole $\ell=220$ from WMAP/Planck lowers the spectral index from ACT. Here ${\cal D}_{\ell=220}$ provides an anchor, which, in a single power law form of primordial spectrum framework, restricts the spectral tilt from being blue~\cite{ACT:2020}. However, this change in tilt, does not imply restoration of consistency as imposing the prior degrades the fit to the ACT data across several multipoles. Since the underlying mechanism for the generation of all the acoustic peaks is the same, the large scales and the small scales spectra should be consistent within and across observations. In Planck~\cite{Planck:2018param} PR3 release, such consistency between parameters estimated from spectra with $2\le\ell\le801$ and $802\le\ell\le2508$ has been shown (see,~\cite{Planck:2016tof} for extensive tests on consistency of {\it Planck} data). Therefore, without bringing the first peak into question, {\it Planck} PR3 and ACT DR4 discrepancy can be specified as small scale inconsistencies between {\it Planck} and ACT. Given the $3\sigma$ significance of this mismatch, three possible sources can be listed:

\begin{enumerate}
    \item Systematic uncertainties: Unknown systematic effects, foreground and calibration uncertainties can give rise to such tensions.
    \item New physics: In this case, given the substantial overlap in temperature data between the two observations, a new physics candidate should be able to independently fit both {\it Planck} and ACT with consistent parameter volume. 
    \item Statistical anisotropy: Since ACT observes only a portion of the full sky observed by Planck, this discrepancy could also indicate statistical anisotropy.     
\end{enumerate}

The statistical chances for disagreement for a random Gaussian field are quite low (we discuss it towards the end of the paper). The error in power spectrum is related to the fraction of sky covered. ACT DR4 covers about 40\% of the sky but that same area is covered by Planck. This extensive overlap makes it difficult to explain the inconsistency with statistical anisotropy. 


In this work, instead of looking for any particular solution to this tension, we try to identify the cosmological scales where {\it Planck} and ACT are different. We make use of two approaches, parametric modelling and non-parametric reconstruction of the primordial power spectrum.

In the parametric modelling, we use an extension of the power law to allow for two different tilts in the spectrum as discussed in~\cite{Hazra:2013broken}. With a kink/break in the power spectrum where the point of transition is allowed to vary, we characterize the transition of the spectral tilt from red to blue.  
Since {\it Planck} and ACT temperature data have significant overlaps, in a joint analysis one has to work with datasets truncated to certain multipoles to avoid statistical bias from double counting. ACT recommends~\cite{ACT:2020} the use of ACT TT spectra after $\ell=1800$ in a joint analysis with Planck. However, in this estimation, the constraint on the tilt will be mainly driven by {\it Planck} due to its access to wide cosmological scales. Therefore, we use data from different combinations of {\it Planck} frequencies and multipoles and combine with ACT avoiding overlap to find the transition scales.  

We use modified Richardson-Lucy~\cite{Richardson:72,Lucy:74} algorithm (MRL)~\cite{HazraMRLWMAP:2013,HazraMRLPlanck:2014} to reconstruct the primordial power spectrum from co-added {\it Planck} and ACT data separately and compare them to understand the disagreement between the two datasets beyond the tilt. Using crossing statistics~\cite{Shafieloo_2012} consistency between different datasets have been tested~\cite{Hazra:2013oqa,Hazra:2014hma,Shafieloo:2016zga}. Here we use MRL to test the consistency between {\it Planck} and ACT in a free form reconstruction. 

The paper is organized as follows. In~\autoref{sec:method} we discuss the parametric and non-parametric analysis details.~\autoref{sec:results} contains the results from both analyses. In~\autoref{sec:summary}, we summarize and comment.
 
\section{Methodology and datasets}~\label{sec:method}
Our analysis is divided into two parts, parametric modelling and non-parametric reconstruction of the primordial power spectrum. In~\autoref{subsec:method-parametric} we discuss the parametric model that we use in this paper, the parameters and their priors. In~\autoref{subsec:data} we discuss the data combinations we use for the parametric analysis. The details on the nested sampling used here are discussed in~\autoref{subsec:sampler}. The non-parametric reconstruction details are discussed in the second part in~\autoref{subsec:method-reconstruction}.

\subsection{Parametric analysis}~\label{subsec:method-parametric}
\subsubsection{Model}
In the baseline model analysis, the power spectrum is expressed by a power law with an amplitude ($A_{s}$) and a spectral tilt ($n_s$) as, 
\begin{equation}~\label{eq:plaw}
    {\cal P}_{S}(k)=A_{s}(k/k_0)^{n_s -1}
\end{equation}

Since~\autoref{eq:plaw} can not capture any change in the slope, it is not suitable to probe the spectral discrepancy between {\it Planck} and ACT. Running of the spectral index allows for a scale dependent tilt but for a change at small scales, the running also imposes a change at large scales. Therefore, in order to have a change in the spectral tilt, without adding too many parameters, we use a broken power law as discussed in~\cite{Hazra:2013broken,Hazra:2014rulingout}. A broken power law refers to a spectrum with two different tilts at two scales. The spectrum ${\cal P}^{\rm broken}_{S}(k)$ can be expressed as,
\begin{equation}~\label{eq:brokenplaw}
    {\cal P}^{\rm broken}_{S}(k)=A_{s}(k=k_{\rm break})\times
\begin{dcases}
    (k/k_{\rm break})^{n_{s1} -1},& \text{if}~k\le k_{\rm break}\\
    (k/k_{\rm break})^{n_{s2} -1},              & \text{if}~k\geq k_{\rm break}
\end{dcases}
\end{equation}
where $n_{s1}$ and $n_{s2}$ are two tilts at scales larger and smaller than $k=k_{\rm break}$ respectively. The amplitude $A_{s}(k=k_{\rm break})$ is defined at the break point $k=k_{\rm break}$.
Here if marginalized $\Delta n_s=n_{s2}-n_{s1}$ posterior rejects the null value ($\Delta n_s=0$), corresponding to~\autoref{eq:plaw}, it will indicate the ruling out of the single power law scenario. 

\subsubsection{Priors}
We use conservative priors on all parameters. We use same priors on the 6 parameters for the baseline model as in CosmoChord~\cite{Handley:CC}. For the two extra parameters, $n_{s2}$ and $\log_{10}k_{\rm break}$ we use the following flat priors in~\autoref{tab:priors}. $\Delta n_s=n_{s2}-n_{s1}$ is obtained as a derived parameter. 

\begin{table}
    \centering
    \begin{tabular}{|c|c|}
    \hline
         Parameters&  Priors \\\hline\hline
         $n_{s2}$ &  [0.8,~1.2] \\\hline
         $\log_{10}{k_{\rm break}}$& [-2.3,~-0.52]  \\
         \hline
    \end{tabular}
    \caption{Priors used for the extra 2 parameters introduced for the broken power law in~\autoref{eq:brokenplaw}. We keep all other priors for the four background and the two power spectrum parameters same as the CosmoChord priors~\cite{gitCosmoChord}. Note that while $n_{s2}$ here denotes the tilt at the smaller scale part of the spectrum, $n_{s1}$ denotes the larger scale part. $k_{\rm break}$ is allowed to vary between $0.005-0.3~{\rm MPc}^{-1}$. This covers the complete overlap between {\it Planck} and ACT and also extends till small scales covered by ACT where CMB is relevant.}
    \label{tab:priors}
\end{table}

\subsection{Datasets and likelihoods}~\label{subsec:data}
We use {\it Planck} data from {\it Planck} PR3 2018 official release~\cite{esaPlanckLegacy} and ACTPOLLite likelihood from ACT DR4 release~\cite{actpollite}. Different {\it Planck} likelihoods that we use in this paper are tabulated in~\autoref{tab:Datasets}. The first column refers to the alias of the likelihood that we will use in this paper for reference. The second column points to the likelihood combinations that the aliases refer to. We should mention that large scale {\it Planck} polarization does not overlap with ACT data (the multipole ranges can be compared from column 3) and is used to constrain the optical depth to reduce the degeneracy with the spectral amplitude. Therefore in the analysis, that attempts to obtain the constraints from ACT only, we also use {\it Planck} large scale polarization likelihood (which we also mention in the figures explicitly). Here plik refers to plik binned likelihood used in {\it Planck} baseline.

We allow all {\it Planck} nuisance parameters containing foreground and calibration parameters to vary as discussed in~\cite{Planck:2018param} and also use the baseline flat or Gaussian priors on these parameters. Since we are using ACTPOL lite likelihood in our analysis, we only allow the default nuisance parameter for polarization efficiency to vary.
\begin{table}
    \centering
    \begin{tabular}{|c|c|c|}
    \hline
         Name& Likelihood &Multipoles \\    \hline    \hline
         P18lowE& EE SimAll~\cite{Planck:2018param}   &2-29   \\ \hline
         P18EE & plik EE high-$\ell$ + P18lowE~\cite{Planck:2018param}&2-1996   \\ \hline
         P18TEEE& plik TEEE high-$\ell$ + P18lowE~\cite{Planck:2018param}&2-1996\\   \hline        
         P18& plik TTTEEE high-$\ell$ & 2-2508 (TT) \\ 
         &  + TT lowL + EE SimAll & 2-1996 (TEEE)~\cite{Planck:2018param} \\ \hline
         
        P18-100GHz ($\ell<650$)& plik TTTEEE high-$\ell$ (only 100 GHz)  & 2-650 \\ 
       &  + TT lowL + EE SimAll~\cite{Planck:2018param} &  \\\hline 
        P18 ($\ell<650$)& plik TTTEEE high-$\ell$ & 2-650 \\ 
       &  + TT lowL + EE SimAll~\cite{Planck:2018param} &  \\\hline 
       
        P18-143-217GHz & plik TTTEEE high-$\ell$  & 2-2508 (TT) \\ 
         & (143 GHz and 217 GHz)~\cite{Planck:2018param}  & 2-1996 (TEEE) \\\hline 

         ACT& ACTPOL lite DR4~\cite{ACT:2020} & 600-4325 (TT) \\
         &  & 350-4325 (TE/EE) \\ \hline
         ACT ($\ell>1800$)& ACTPOL lite DR4~\cite{ACT:2020}& 1800-4325 (TT) \\
         &  & 350-4325 (TE/EE) \\ \hline
         
    \end{tabular}
    \caption{Likelihoods and multipole ranges used in this article. Hereafter, different combinations of likelihoods will be referred with the names mentioned in the first column. We select different frequency channels in order to understand the scale of the discrepancy between {\it Planck} and ACT which is elaborated in the text.}
    \label{tab:Datasets}
\end{table}

\subsection{Sampling details}~\label{subsec:sampler}
We expect multimodal posteriors for the primordial power spectrum parameters emerging from different preferences for tilts from different observations covering different scales. Therefore, instead of Markov Chain Monte Carlo sampling, we use nested sampling. We use PolyChord~\cite{Handley:CC} sampler for this purpose. We use 1024 live points in all the analyses and use a convergence criteria of 0.01 in the logarithm of evidence. 

\subsection{Non-parametric reconstruction}~\label{subsec:method-reconstruction}
In order to understand the difference between {\it Planck} PR3 and ACT DR4 at higher orders than the tilt, we perform a non-parametric reconstruction. We use the MRL method developed in~\cite{HazraMRLWMAP:2013,HazraMRLPlanck:2014}. Being non-parametric, the method has the advantage of reconstructing all features present in the data that the baseline can not address. Simultaneously it extracts the wide and the localized features. Therefore, it is possible to highlight the scales of disagreement directly. 
However, in non-parametric methods, it is difficult to disentangle the physical effects from statistical fluctuations and noise. Staring from an initial guess (${\cal P}^0_k$) of the primordial spectrum (the result however, is independent of the initial guess~\cite{ShafielooIRL:2004}) MRL reconstructs primordial spectrum in an iterative way as provided in~\autoref{eq:mrl}. 
\begin{eqnarray}~\label{eq:mrl}
\frac{{{\cal P}_{k}^{(i+1)}}-{{\cal P}_{k}^{(i)}}}{{\cal P}_{k}^{(i)}}&=&\sum_{\ell=\ell_{\rm min}}^{\ell_{\rm cutoff}}
{\widetilde{G}}_{\ell k}\Biggl\{\l(\frac{{\cl^{\rm {D'}}}-\cl^{{\rm T}(i)}}{\cl^{{\rm T}(i)}}\r)~\tanh^{2}
\l[Q_{\ell} (\cl^{\rm {D'}}-\cl^{{\rm T}(i)})\r]\Biggr\}_{\rm unbinned}\nn\\
&+&\sum_{\ell=\ell_{\rm cutoff}}^{\ell_{\rm max}}
{\widetilde{G'}}_{\ell k}\Biggl\{\l(\frac{\cl^{\rm {D'}}-\cl^{{\rm T}(i)}}{\cl^{{\rm T}(i)}}\r)~\tanh^{2}
\l[\frac{\cl^{\rm D'}-\cl^{{\rm T}(i)}}{{\sigma_{\ell}^{\rm D}}}\r]^{2}\Biggr\}_{\rm binned}
\end{eqnarray}

Here ${\cal P}_{k}^{(i+1)}$ at $i+1$'th iteration is reconstructed as a modification to the $i$'th iteration spectrum. $\cl^{\rm {D'}}$ represents the CMB angular power spectrum, where we have subtracted the lensing effect (obtained from the best fit baseline model) from the co-added angular power spectrum data. $\cl^{{\rm T}(i)}$ represents the theoretical angular power spectrum from $i$'th iteration. ${\widetilde{G}}_{\ell k},{\widetilde{G'}}_{\ell k}$ are the normalized transfer functions in unbinned and binned cases, respectively (note that the unlensed angular power spectrum and the primordial spectrum are related by a convolution ${\cal C}_{\ell}=\sum_i G_{\ell k(i)} P_{k(i)}$). $Q_{\ell},\sigma_{\ell}$ represent the error in the data and are defined in~\cite{HazraMRLPlanck:2014}. This method allows us to reconstruct the primordial spectrum both from unbinned and binned data, which is particularly useful in this context as ACT bandpowers are binned ($\ell_{\rm cutoff}$ represent the multipole after which we switch to reconstruction from binned data). In this work we perform two reconstructions.

\begin{enumerate}

    \item Plik co-added TT angular power spectrum from used~\cite{PLA}: We use unbinned data from PR3. The background parameters are fixed at PR3 best fit obtained from Plik TTTEEE + low-l+ Simall baseline bestfit. This reconstruction apart from providing a reference for the {\it Planck}+ACT reconstruction, also provides a consistency check. 
    
    \item {\it Planck}+ACT reconstruction: Here {\it Planck} TT data is used till $\ell=574$ and ACT DR4 TT data are used afterwards (here $\ell_{\rm cutoff}=574$ since the ACTPOL temperature power spectrum bin starts from $\ell=575$, with central multipole of the first bin being $\ell=600$). Here too background parameters are fixed at PR3 best fit to consistently compare our results from the {\it Planck} reconstruction. While we are not interested in large scale features, the use of {\it Planck} TT data till $\ell=574$ ensure realistic reconstructions of the primordial and angular spectrum at large scales which contributes to smaller scale reconstruction owing to the distribution of transfer function over large range of multipoles. 
\end{enumerate}

As discussed in~\cite{HazraMRLPlanck:2014}, with higher iterations, short-length correlated features, noise gets imprinted in the reconstruction. In this work, since we are mainly interested in understanding broad discrepancies between the two surveys, we restrict ourselves to fewer iterations. We do not use EE and TE datasets for reconstructions as we do not have significant overlap between {\it Planck} and ACT data and therefore we do not expect the reconstructions to highlight any discrepancies. Errorbars on the reconstructed spectra are obtained with reconstructed samples from 1000 realizations of the data drawn from the original data. From the reconstructed samples we also numerically evaluate the effective spectral tilt following $n_s^{\rm eff}=1+\frac{d\ln {\cal P}(k)}{d\ln k}$.
 
\section{Results}~\label{sec:results}

\subsection{Results from the parametric model analysis}

We provide the constraints obtained from different dataset combinations. In~\autoref{tab:comparison}, for different datasets, we tabulate the improvement in fit obtained with the broken power law compared to the baseline and the Bayes' factor. 
\begin{figure}[!htb]
\includegraphics[width=0.8\textwidth]{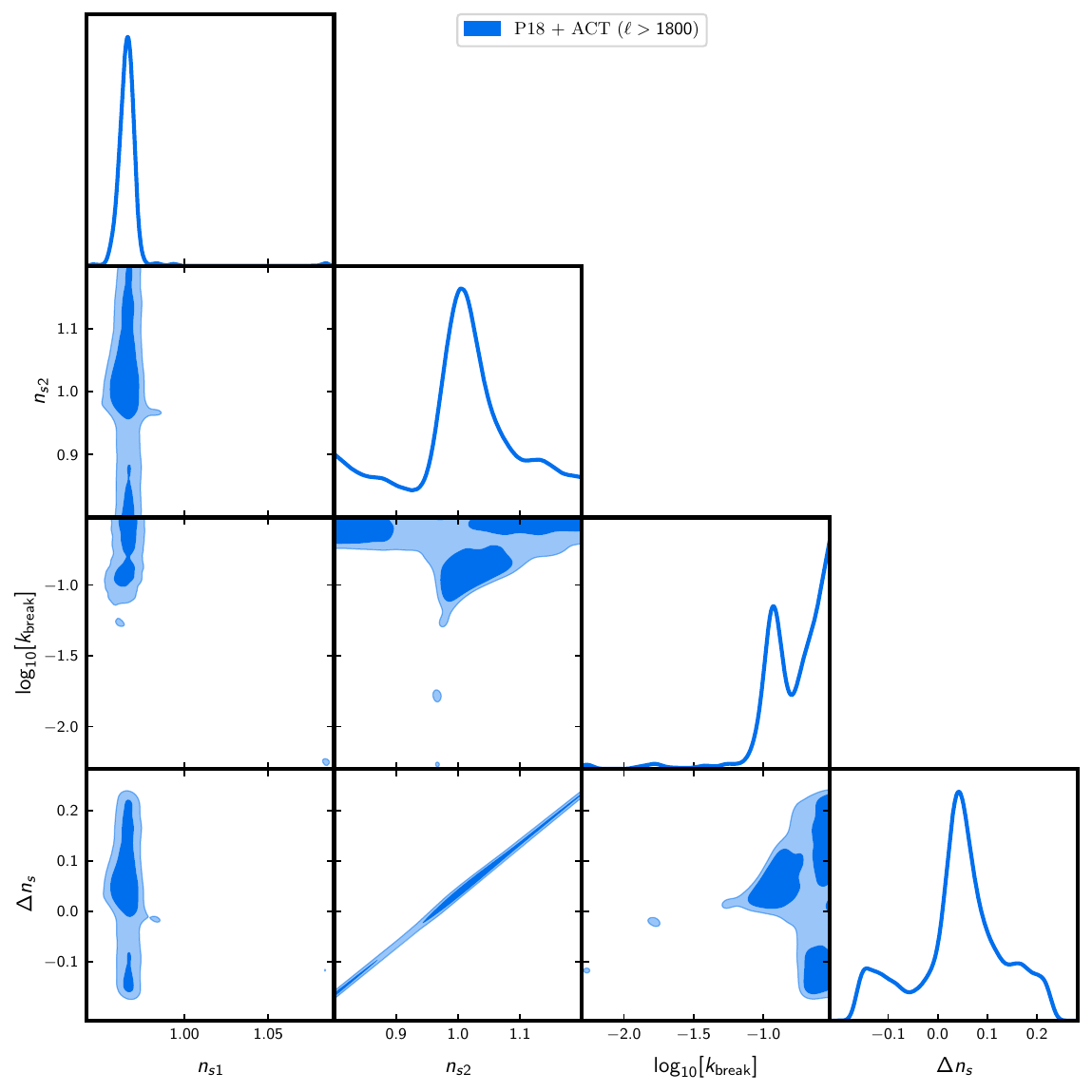}

\includegraphics[width=0.49\textwidth]{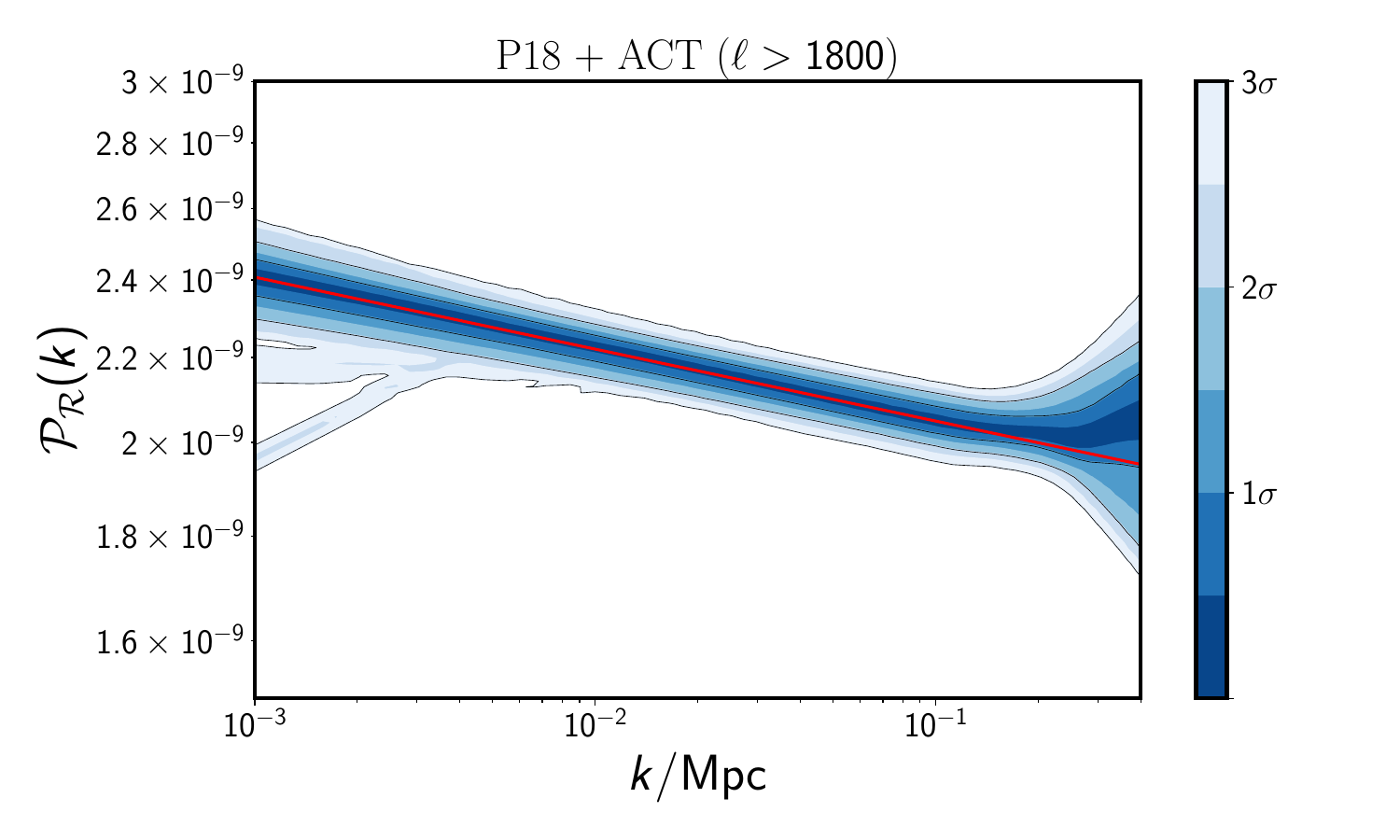}
\includegraphics[width=0.49\textwidth]{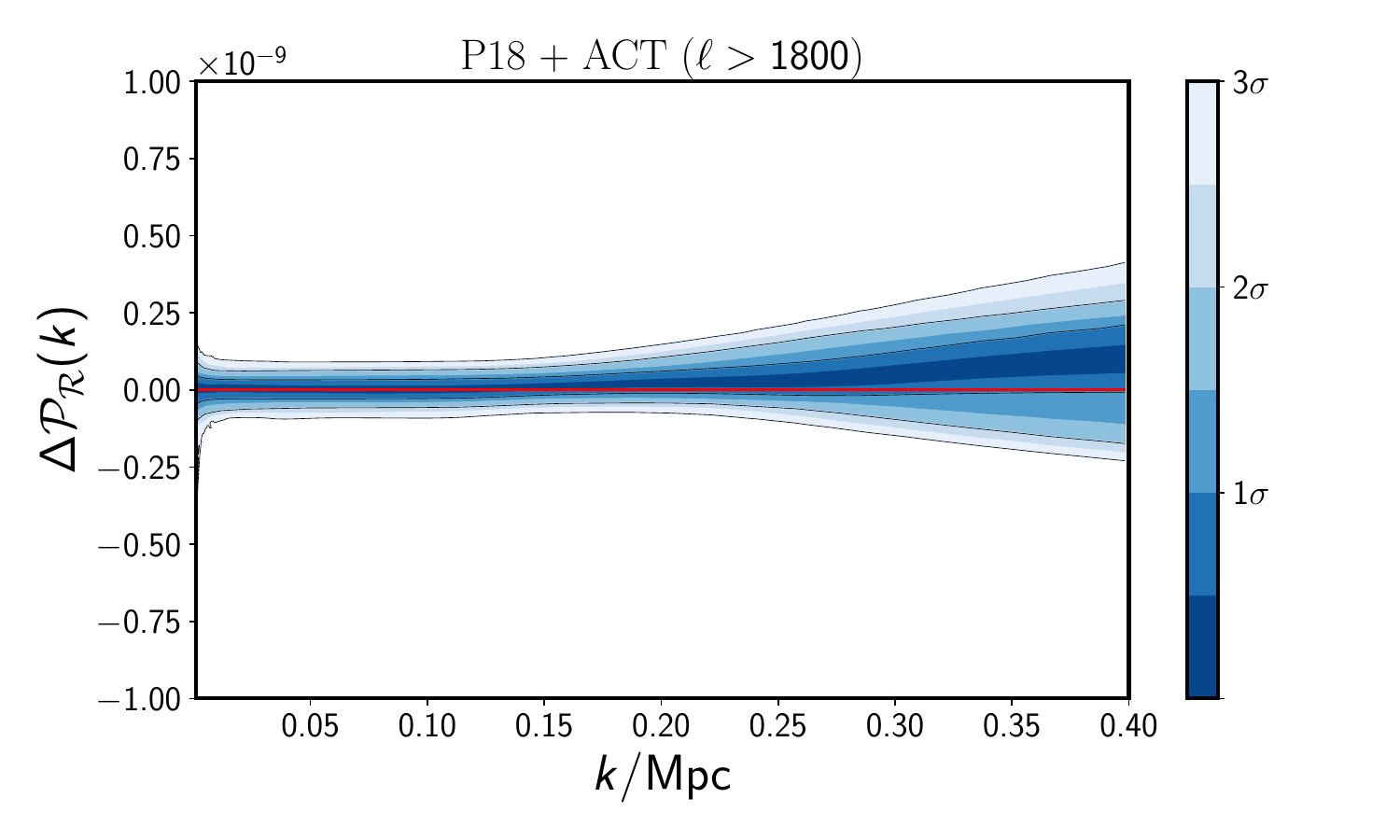}
\caption{Results from P18+ACT ($\ell>1800$) analysis. Top panel contains the parameter posterior triangle plot while the bottom plots contain the power spectrum and the residual power spectrum posterior.} \label{Fig:P18-ACT}
\end{figure}

\begin{figure}[!htb]
\includegraphics[width=0.8\textwidth]{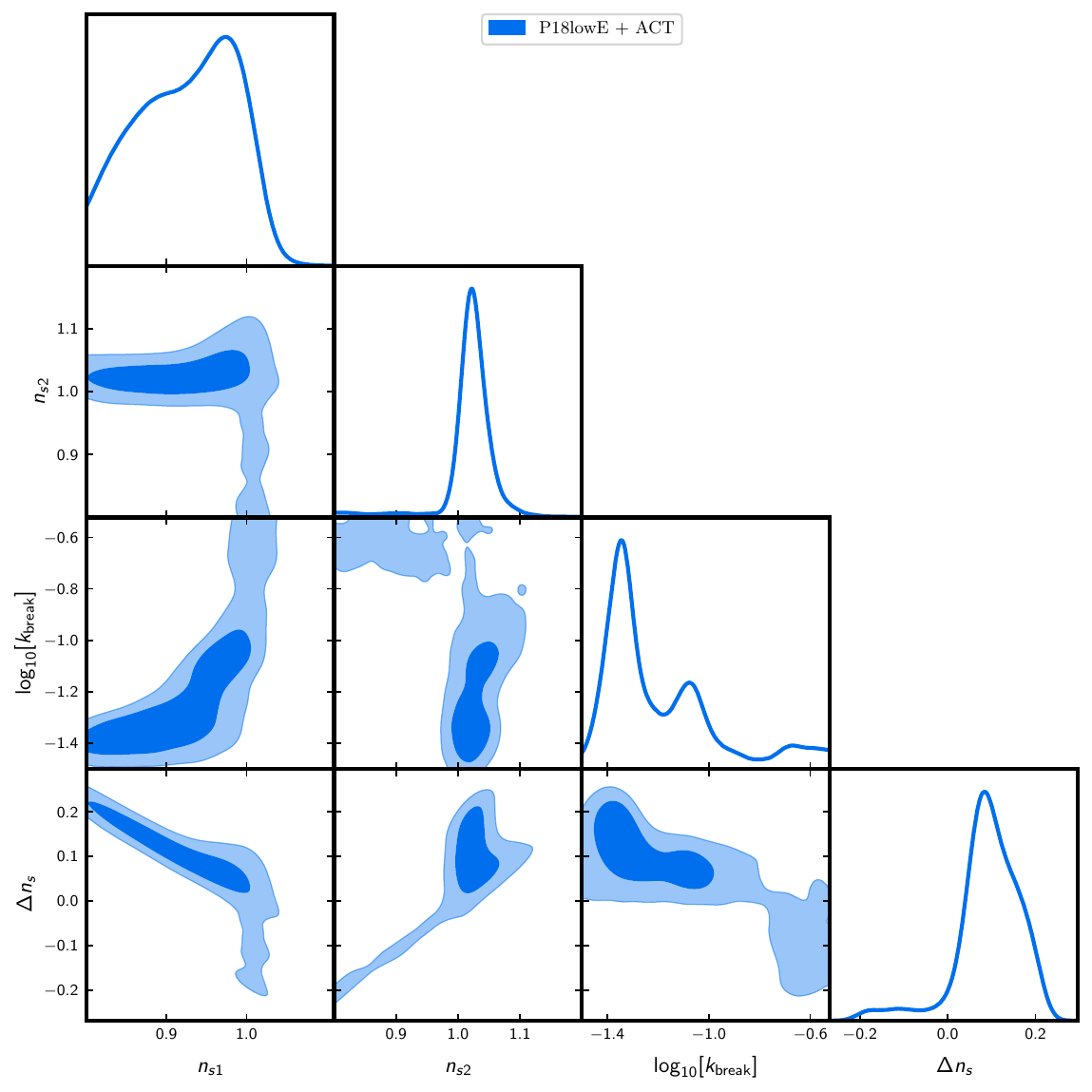}

\includegraphics[width=0.49\textwidth]{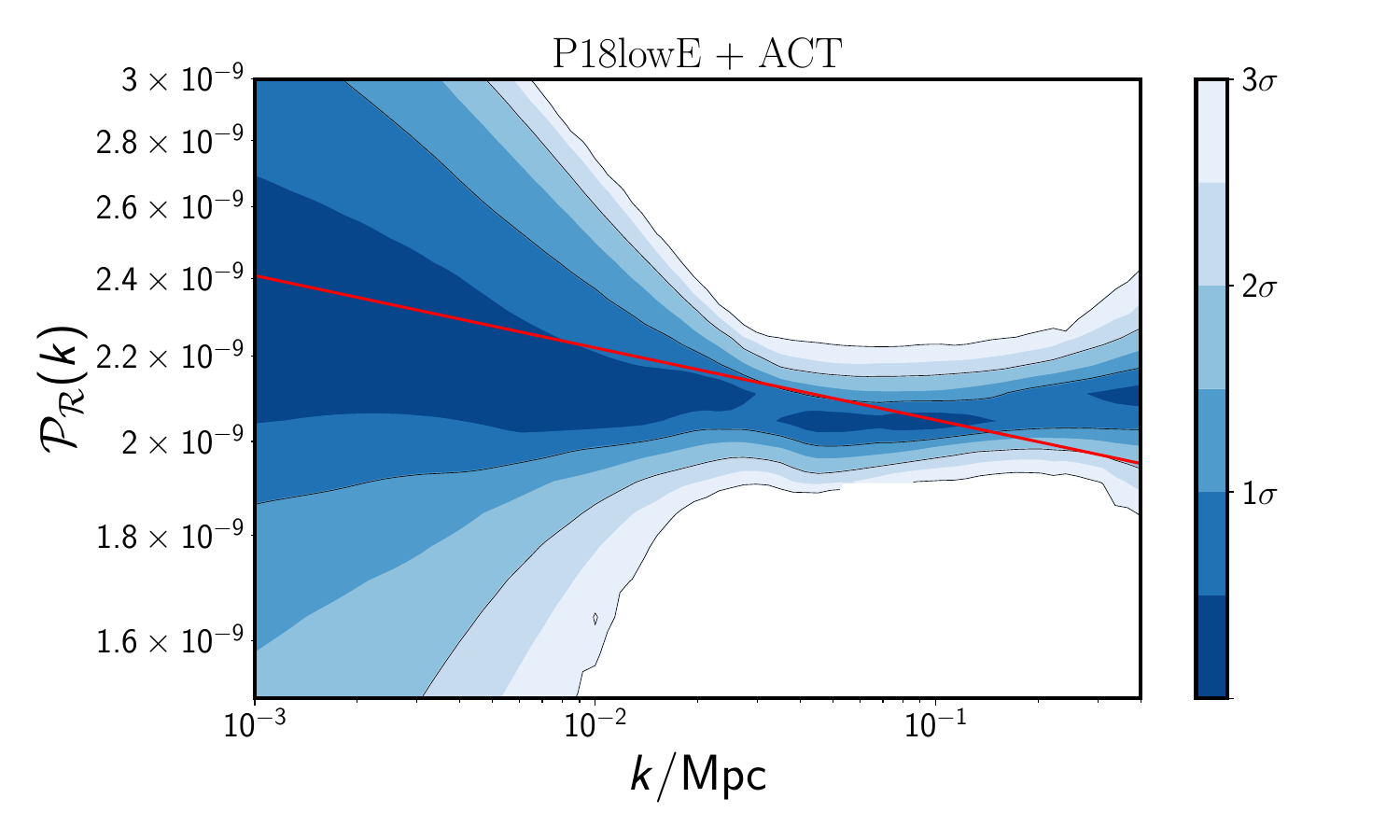}
\includegraphics[width=0.49\textwidth]{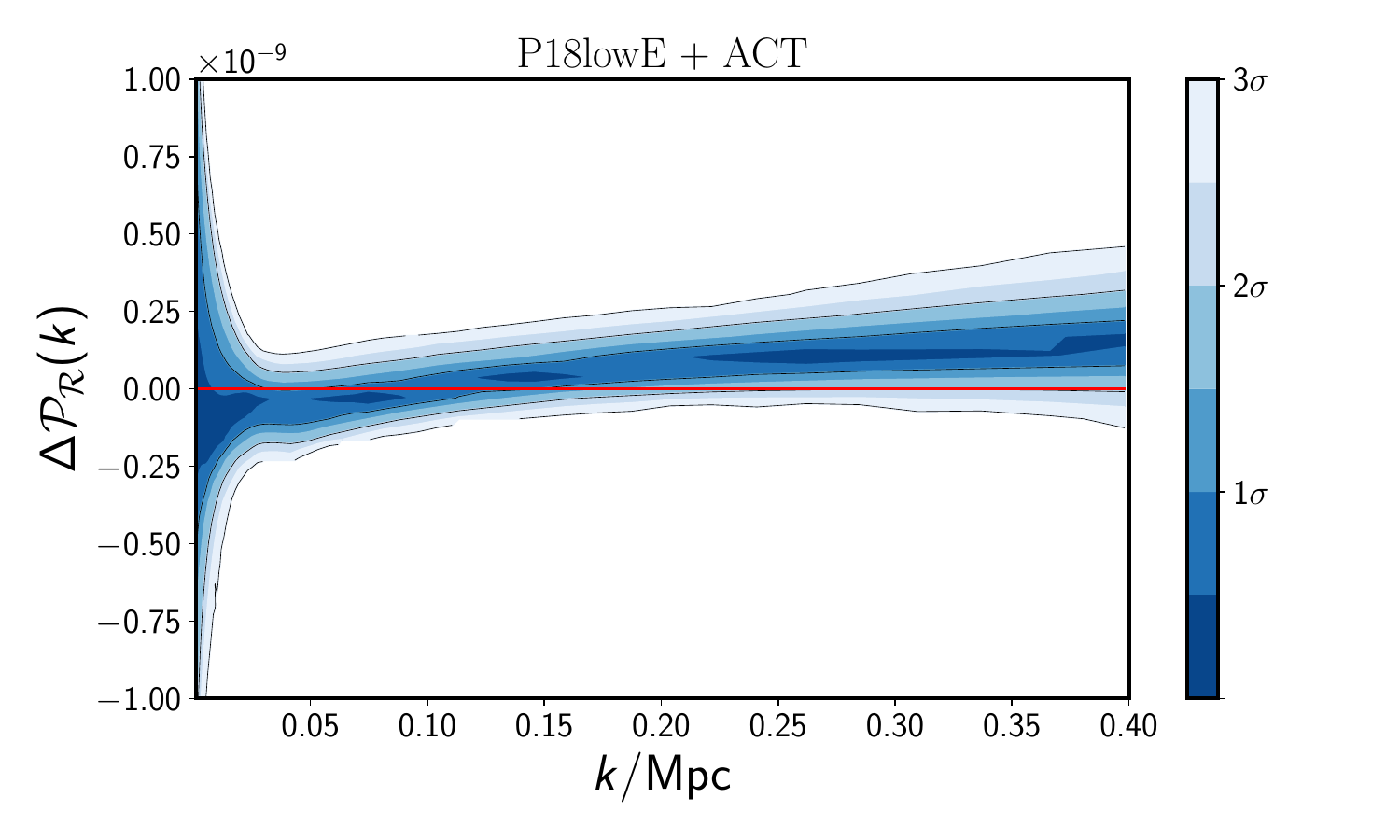}
\caption{Results from P18lowE+ACT analysis. Top panel contains the parameter posterior triangle plot while the bottom plots contain the power spectrum and the residual power spectrum posterior. Here the P18lowE is used to constrain the optical depth, which in turn reduces the degeneracy with spectral amplitude.} \label{Fig:ACT}
\end{figure}

\begin{figure}[!htb]
\includegraphics[width=0.8\textwidth]{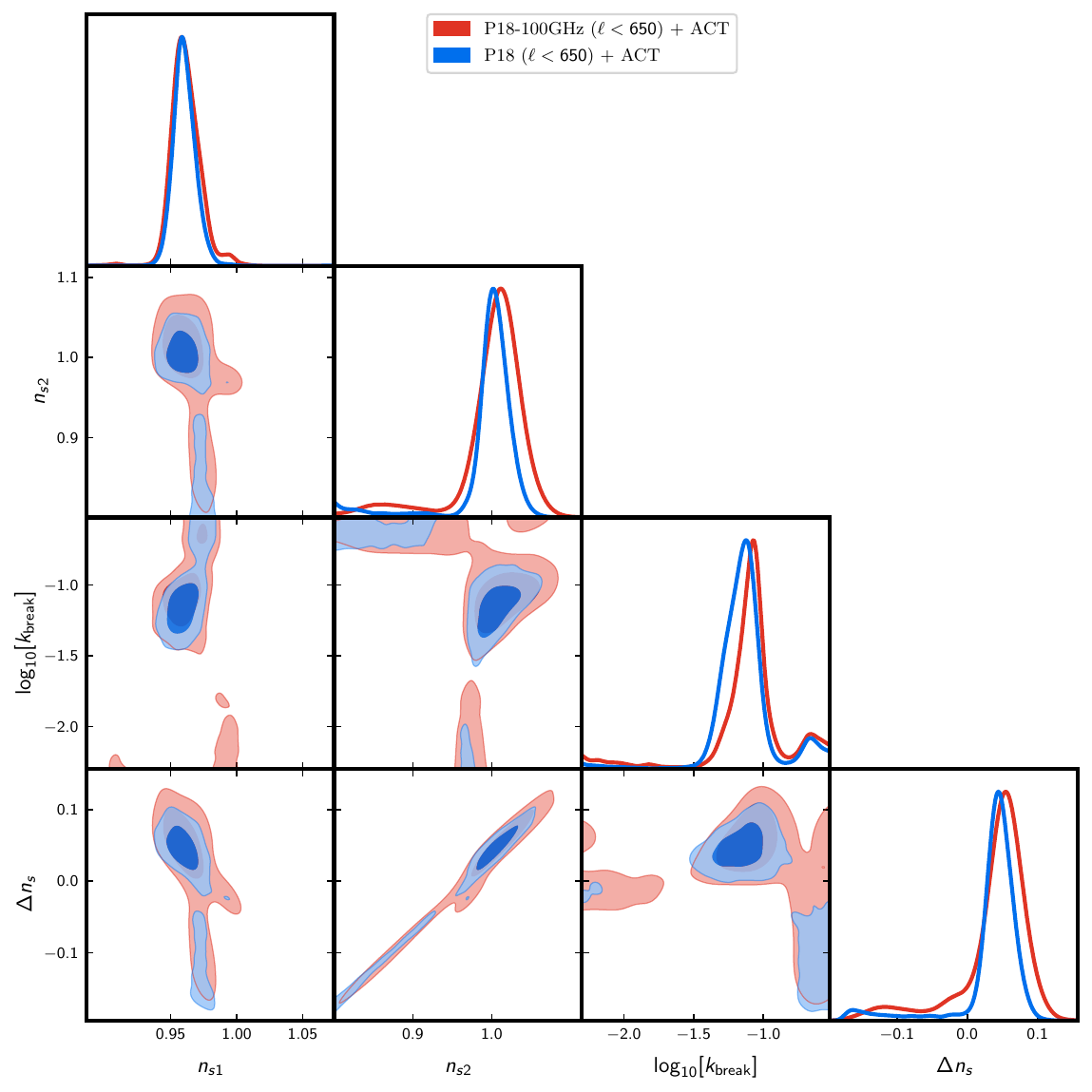}

\includegraphics[width=0.49\textwidth]{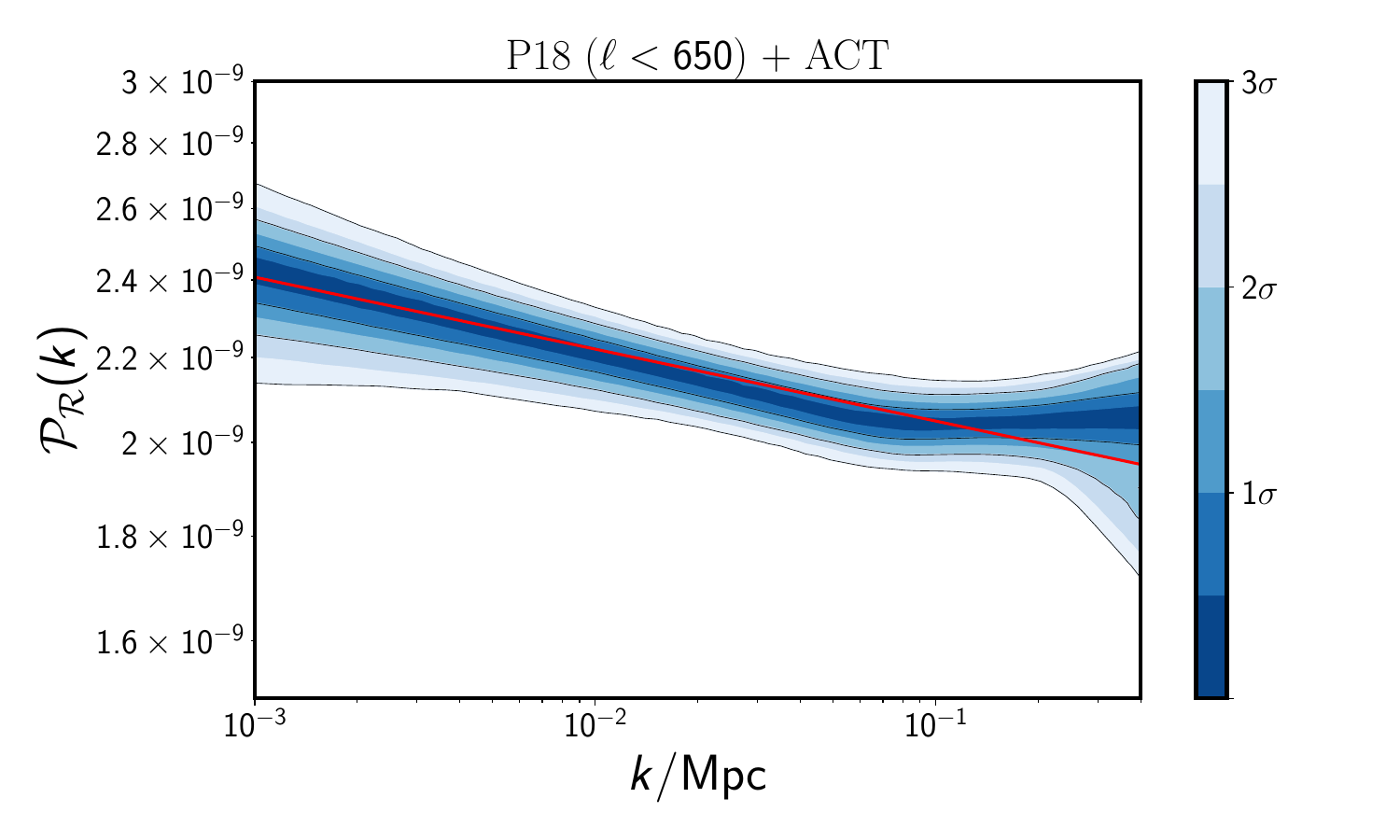}
\includegraphics[width=0.49\textwidth]{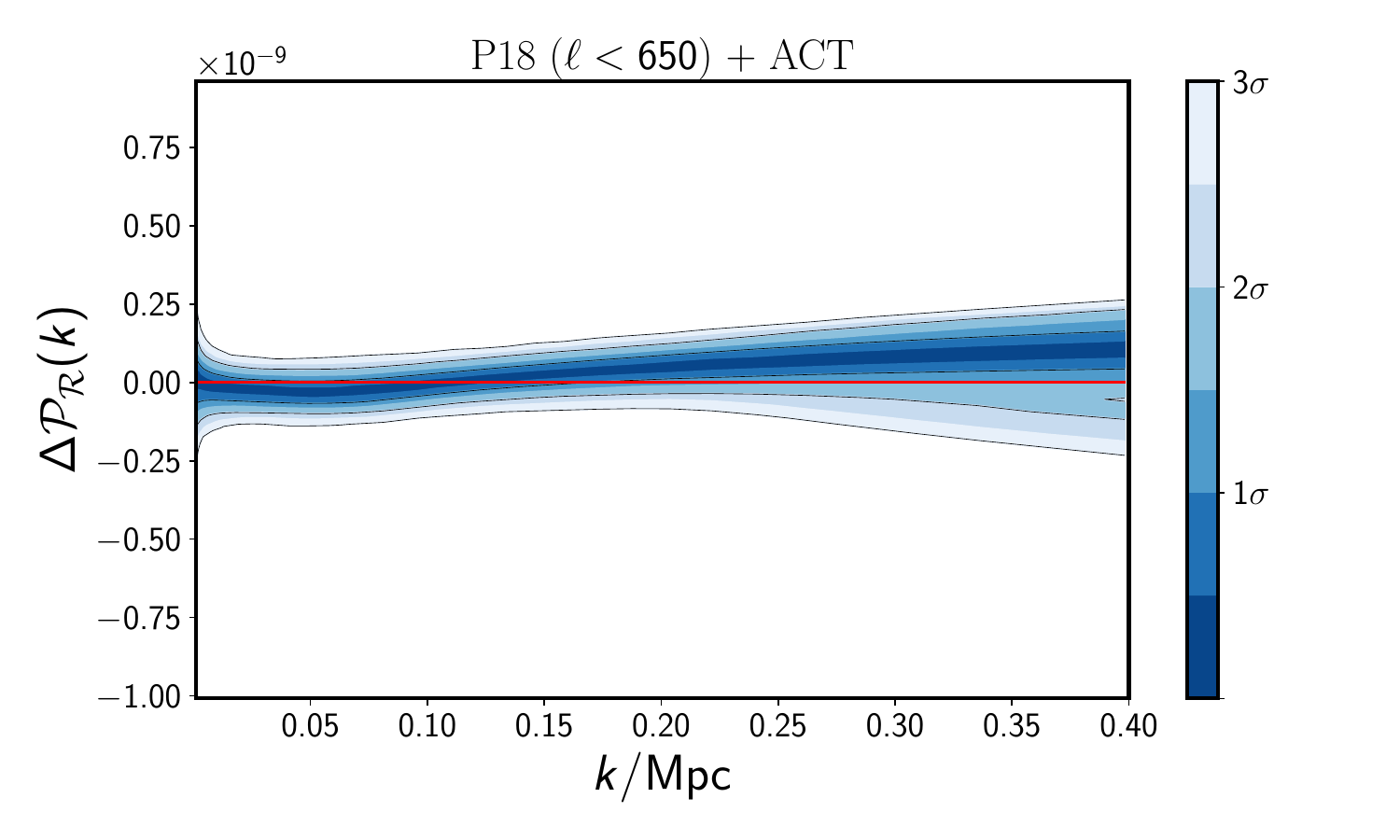}
\caption{Results from P18-100GHz ($\ell<650$)+ACT and P18 ($\ell<650$)+ACT analysis. Top panel contains the parameter posterior triangle plot while the bottom plots contain the power spectrum and the residual power spectrum posterior. While 100GHz spectrum from {\it Planck} till $\ell=650$ and ACT spectrum afterwards provide supports for the change in the spectral tilt, the other spectra from {\it Planck} increases the statistical significance to 3$\sigma$.} \label{Fig:P18100-ell650-ACT}
\end{figure}
\begin{figure}[!htb]
\includegraphics[width=0.8\textwidth]{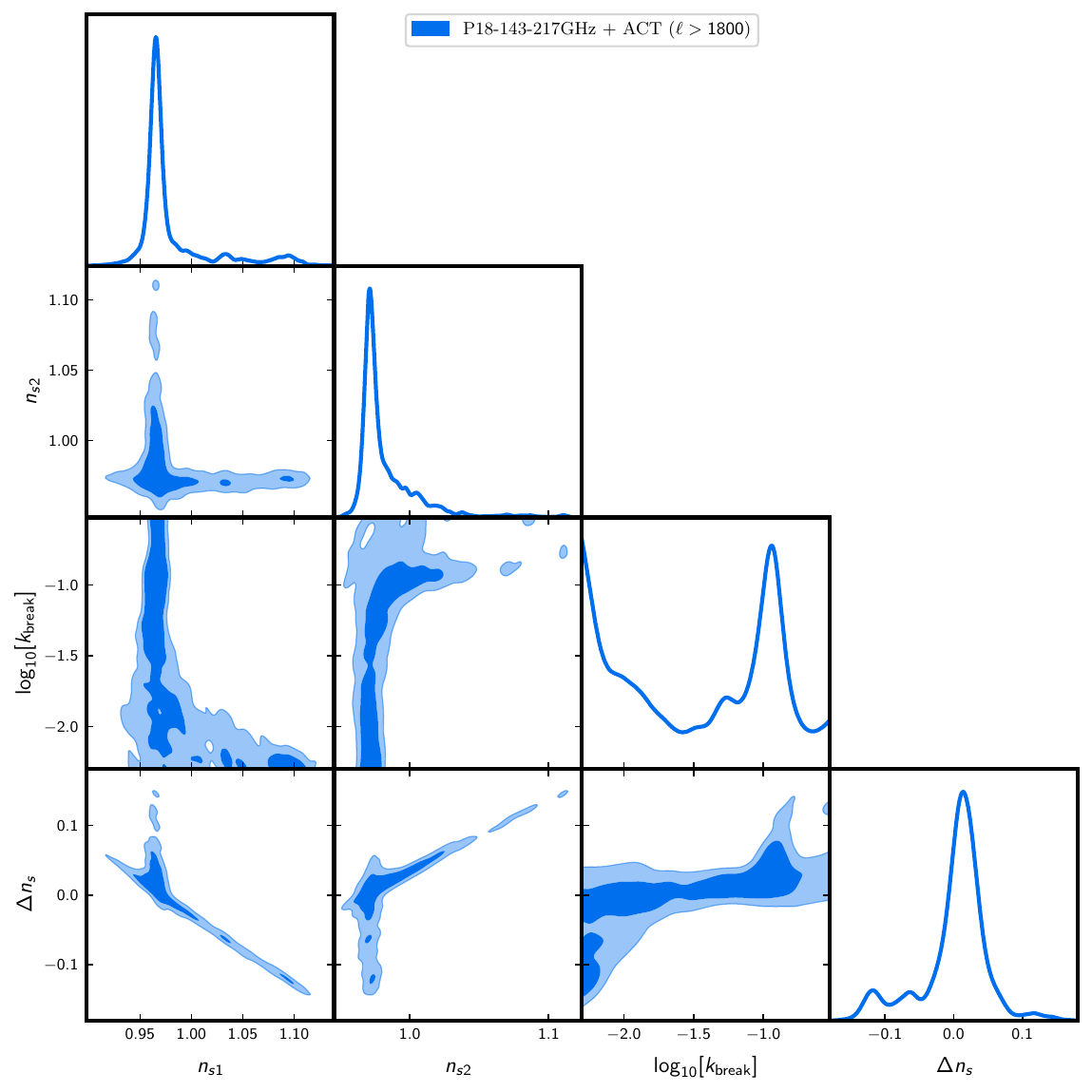}

\includegraphics[width=0.49\textwidth]{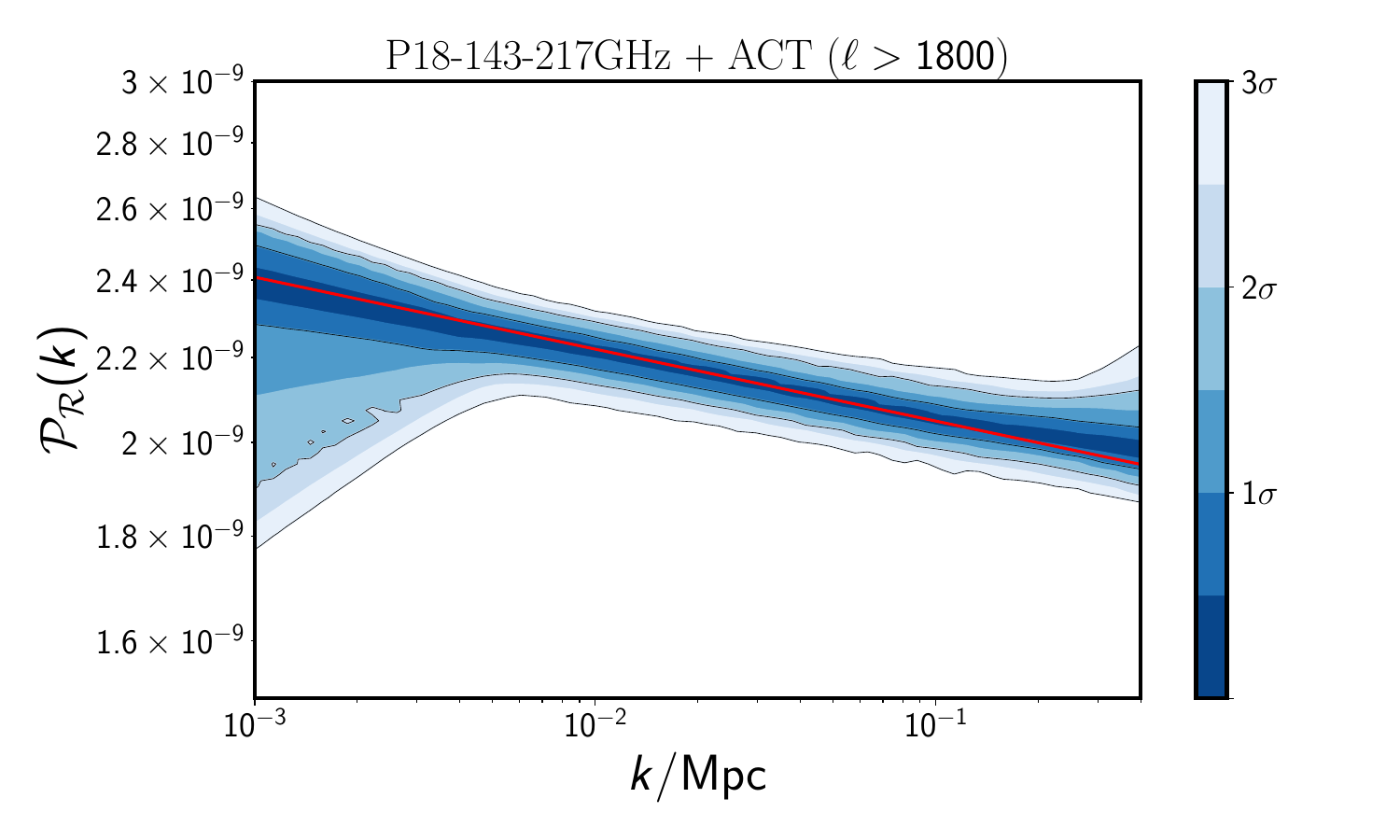}
\includegraphics[width=0.49\textwidth]{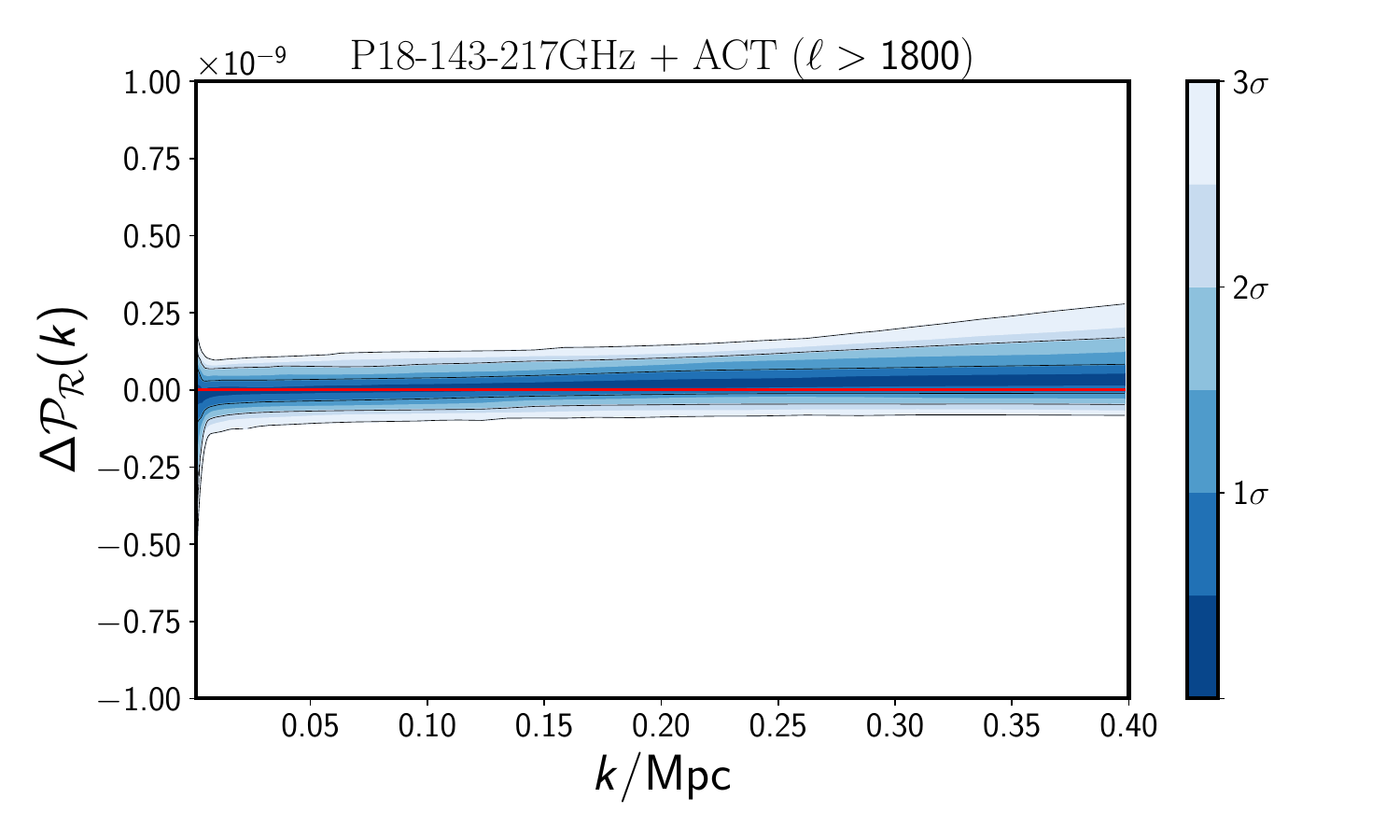}
\caption{Results from P18-143-217GHz+ACT ($\ell>1800$) analysis. Top panel contains the parameter posterior triangle plot while the bottom plots contain the power spectrum and the residual power spectrum posterior.} \label{Fig:P18143217-ACT}
\end{figure}
\begin{figure}[!htb]
\includegraphics[width=0.8\textwidth]{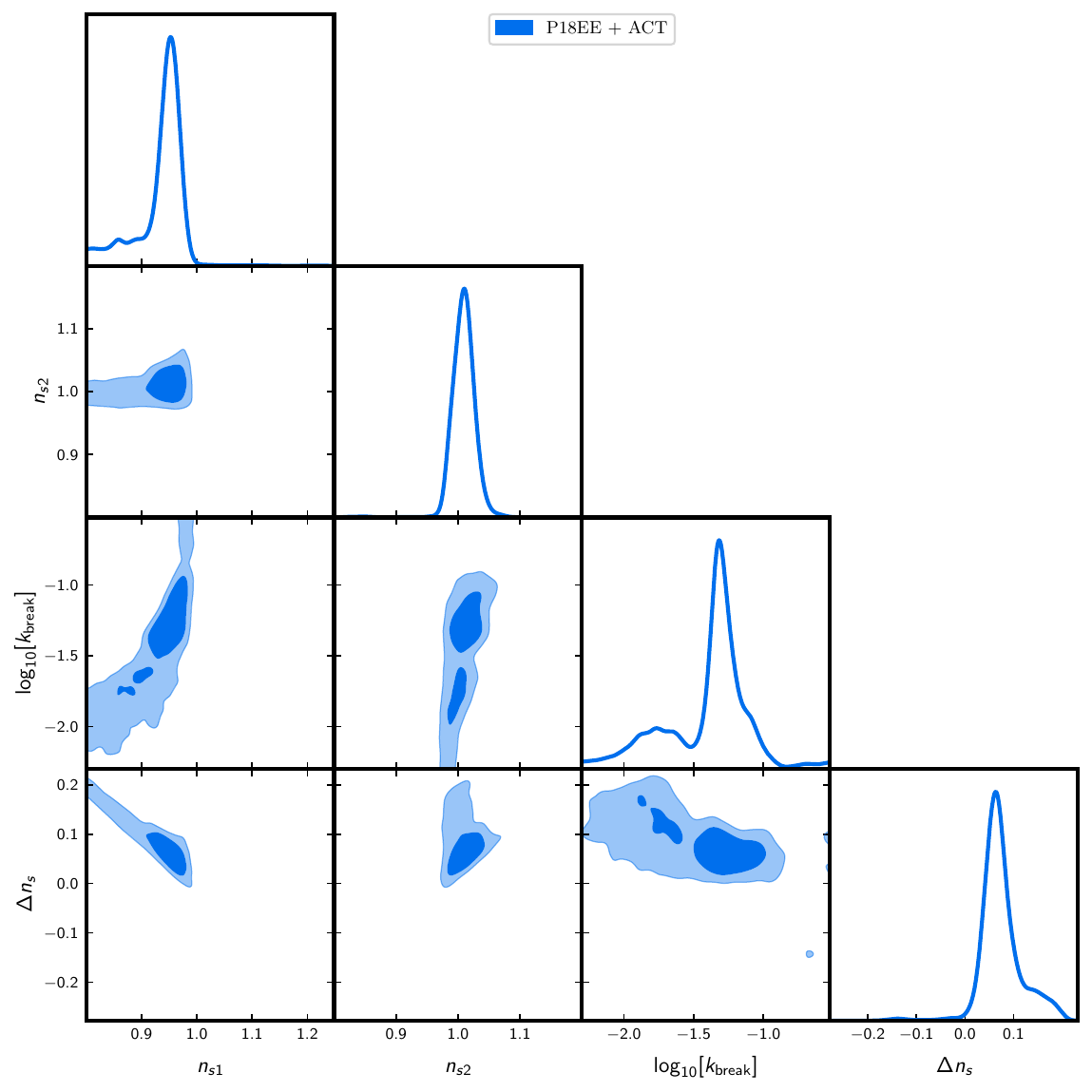}

\includegraphics[width=0.49\textwidth]{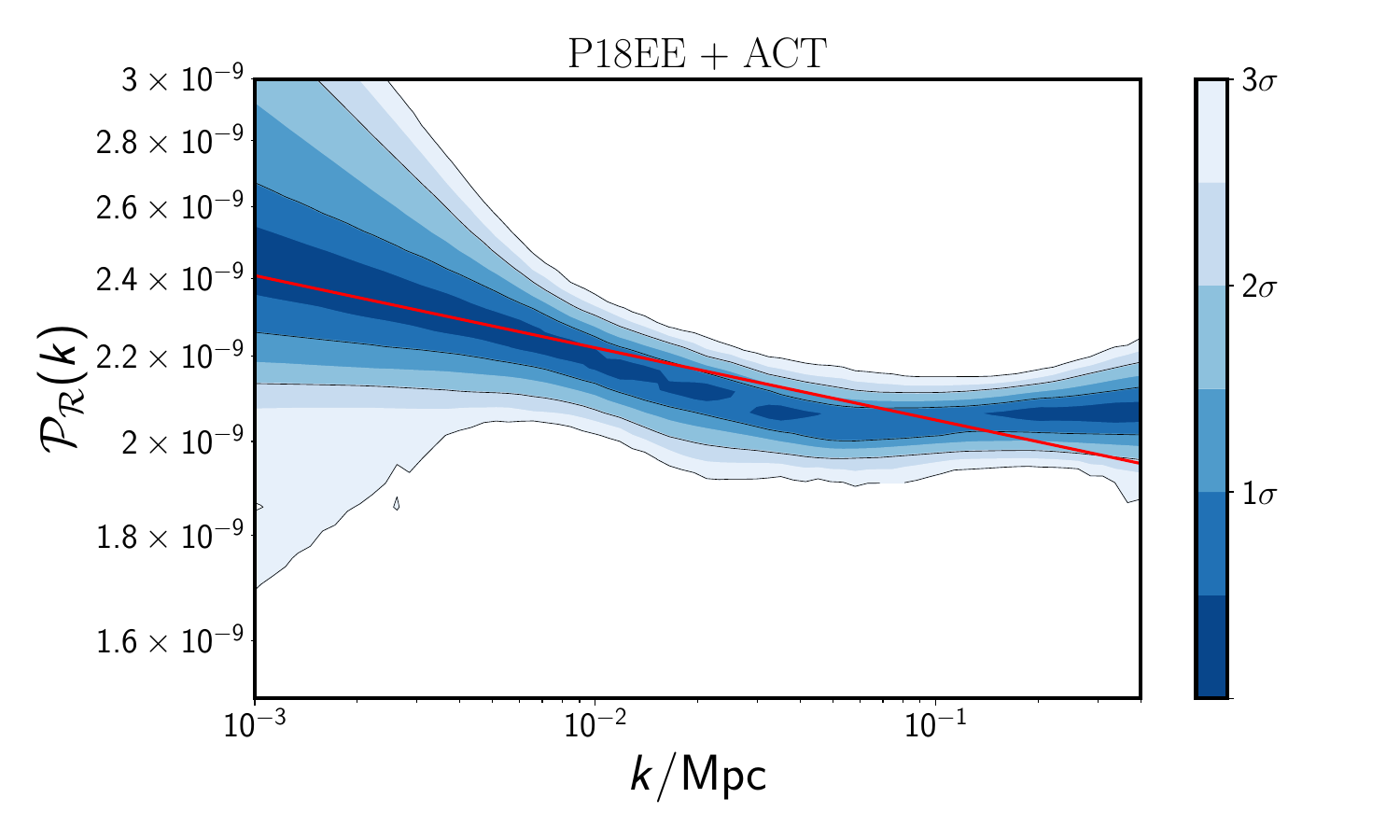}
\includegraphics[width=0.49\textwidth]{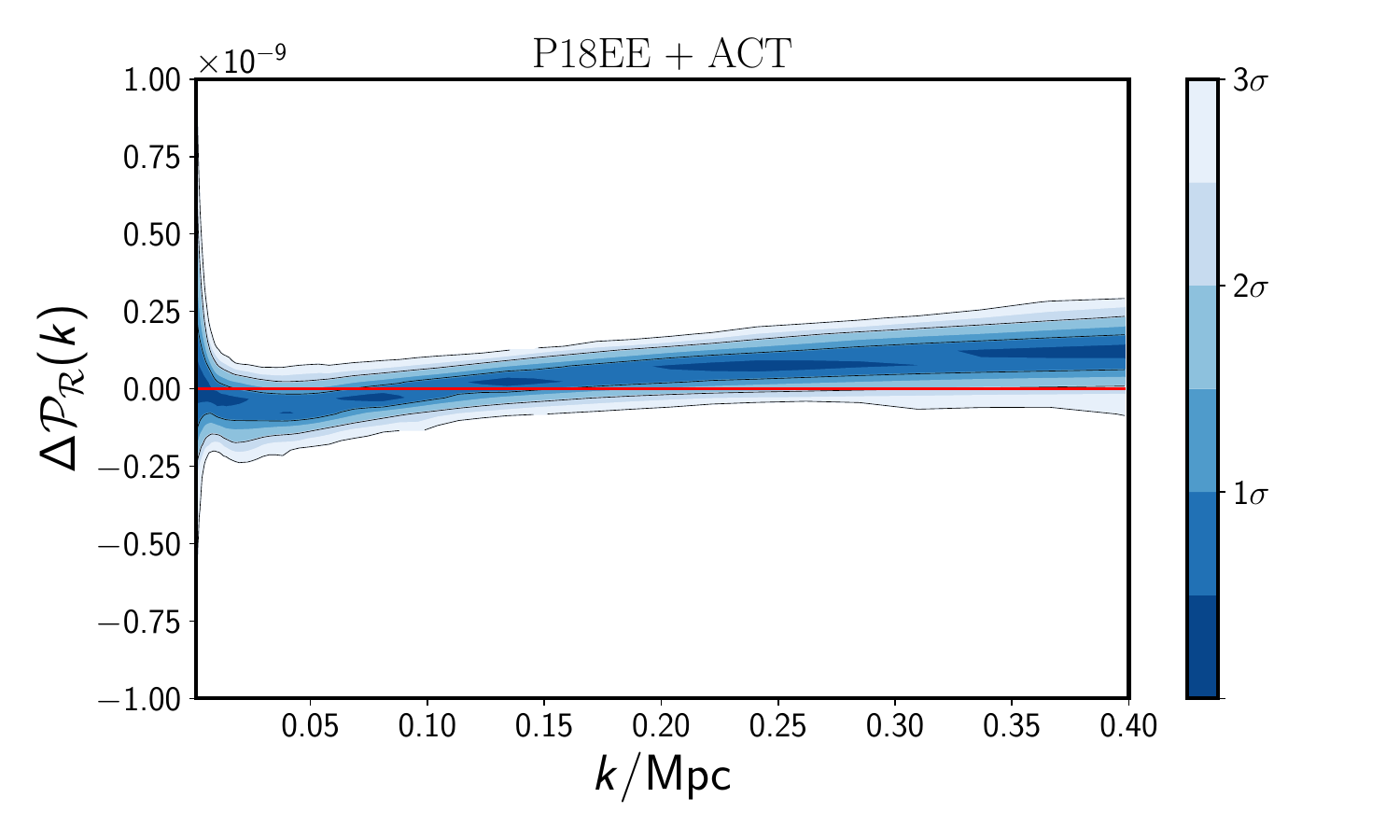}
\caption{Results from P18EE+ACT analysis. Top panel contains the parameter posterior triangle plot while the bottom plots contain the power spectrum and the residual power spectrum posterior.} \label{Fig:P18EE-ACT}
\end{figure}

\begin{figure}[!htb]
\includegraphics[width=0.8\textwidth]{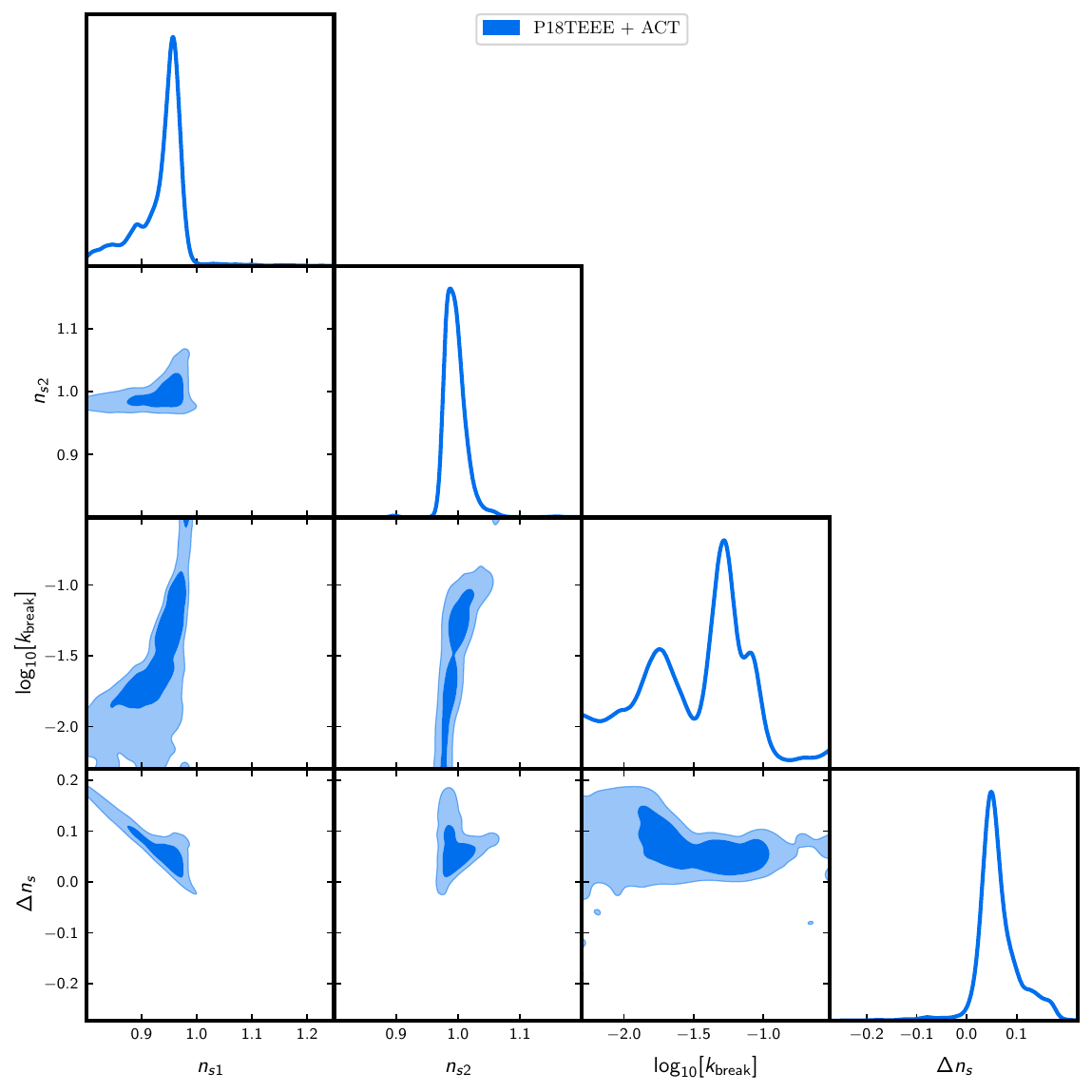}

\includegraphics[width=0.49\textwidth]{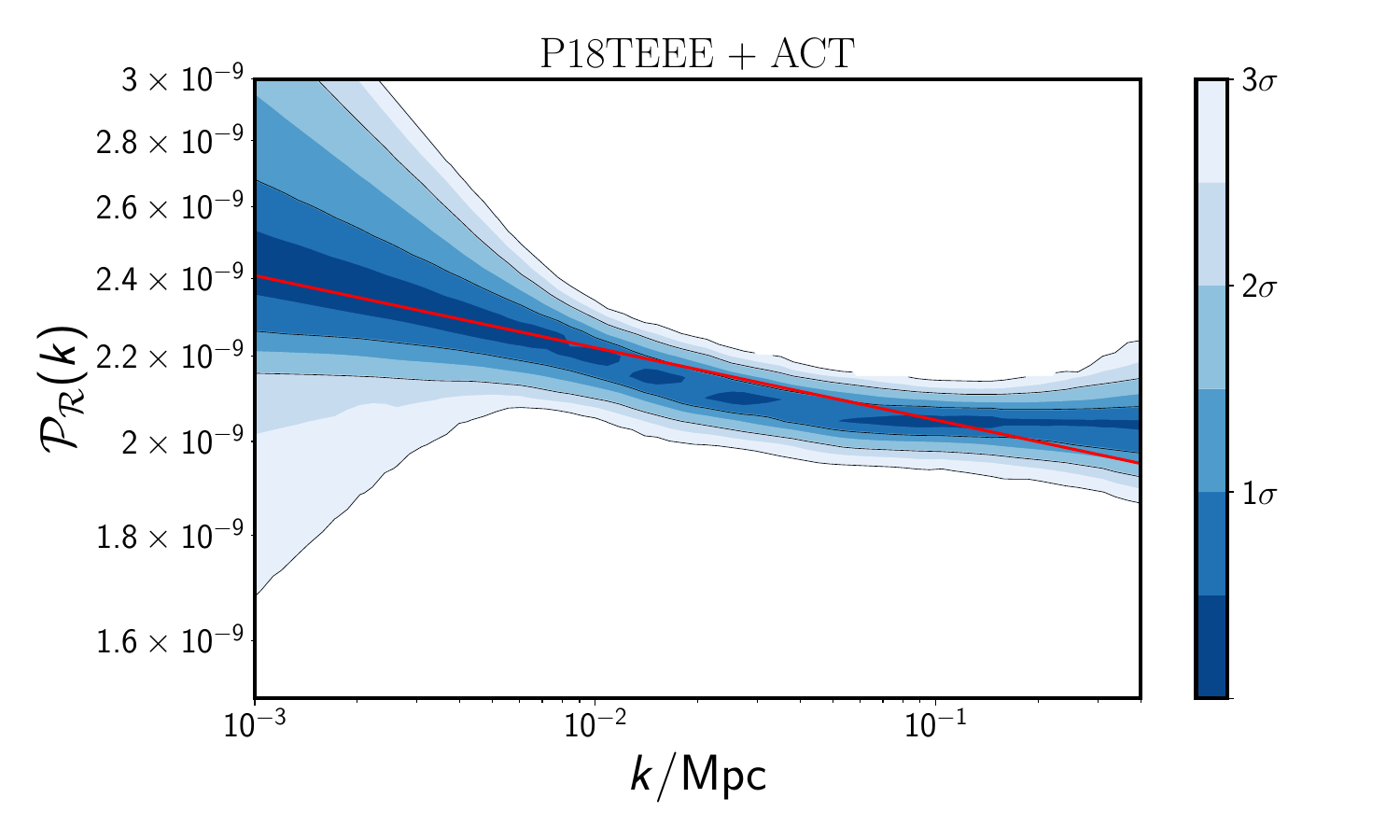}
\includegraphics[width=0.49\textwidth]{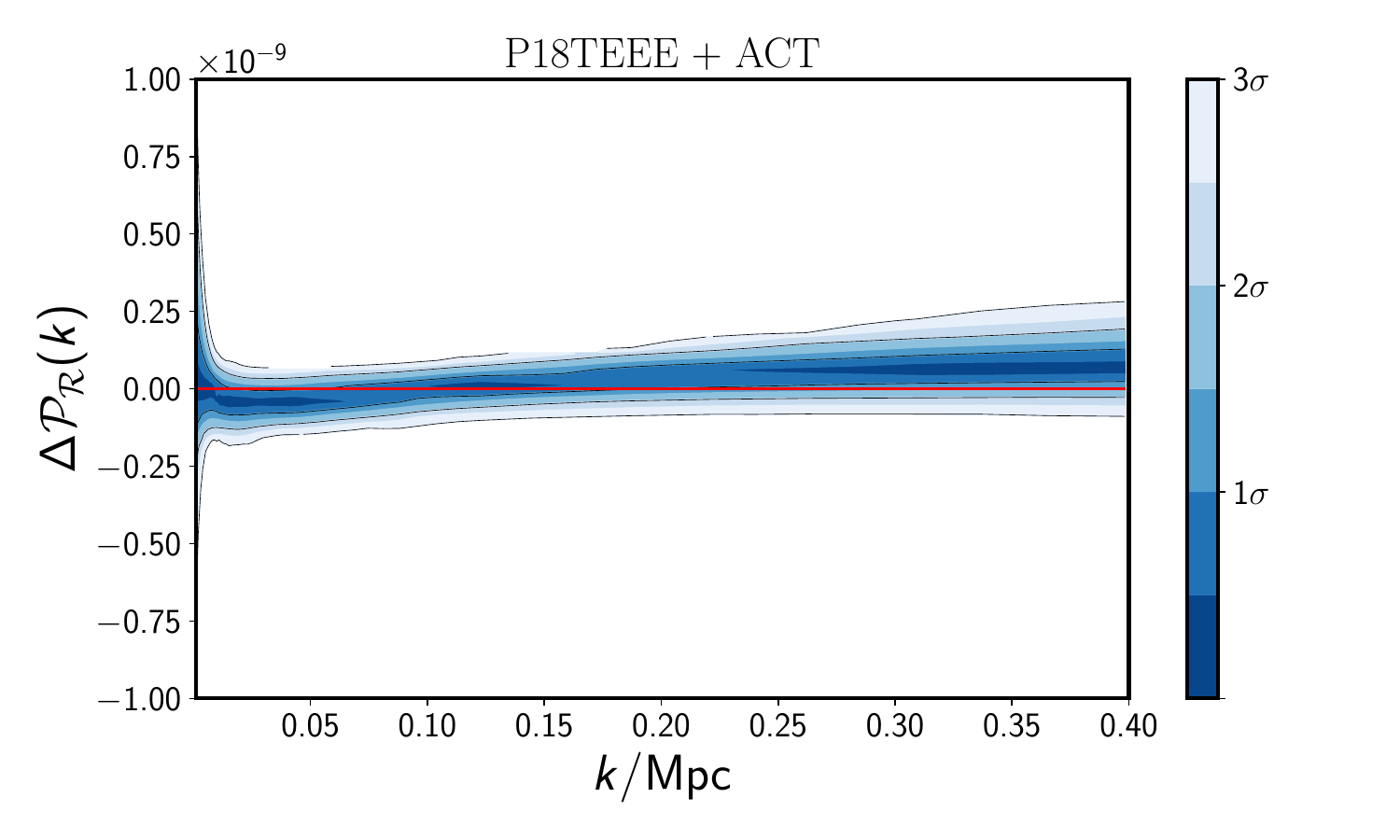}
\caption{Results from P18TEEE+ACT analysis. Top panel contains the parameter posterior triangle plot while the bottom plots contain the power spectrum and the residual power spectrum posterior.} \label{Fig:P18TEEE-ACT}
\end{figure}

\begin{enumerate}
    \item {\bf P18+ACT ($\ell>1800$)}: To begin with we present the results for P18+ACT ($\ell>1800$) in~\autoref{Fig:P18-ACT}. This is the recommended combination by the ACT collaboration~\cite{ACT:2020}. The {\it Planck} complete temperature and polarization data is used while ACT TT data is used after multipole 1800. We find $n_{s1}$ posterior is distributed only in red tilt region indicating the preference from Planck. $n_{s2}$, however is distributed on either side of 1. For $\log_{10} [k_{\rm break}]<-1.24$, where the break in the spectrum occurs at scales $0.056/{\rm Mpc}$, the preferred $n_{s1}$ is bluer than {\it Planck} best fit. The functional posterior of the primordial spectrum and its residual {\it w.r.t.} the {\it Planck} best fit are plotted in the bottom panels. At small scales, $k\sim0.15~{\rm Mpc}^{-1}$, we do not find any strong preference of blue tilt driven by ACT.
    
    \item {\bf P18lowE+ACT }: \autoref{Fig:ACT} plots the constraints from ACT only. The {\it Planck} low-multipole EE likelihood is used to constrain the optical depth so that the scalar spectral amplitude can be better constrained. As expected, without the large scale CV limited temperature data from Planck, $n_{s1}$ can not be constrained well. Here the positive correlation between $\log_{10} [k_{\rm break}]$ and $n_{s1}$ clearly shows that with smaller scales, the large scale power spectrum index turns blue. Similarly $n_{s1}$ and $n_{s2}$ 2D confidence regions also highlight, while $n_{s1}$ is not strongly constrained (dependent on the scale $k_{\rm break}$), $n_{s2}>1$ is preferred. $\log_{10} [k_{\rm break}]-n_{s2}$ contour has two disconnected islands. At $\log_{10} [k_{\rm break}]>-0.8$ ($k_{\rm break}>0.158/{\rm Mpc}$) a blue tilt is ruled out. 
    The correlation between $\Delta n_s$ and $\log_{10} [k_{\rm break}]$ reveals that smaller than this scale, ACT data prefers a $n_{s2}<n_{s1}$, which, in turn, implies that the larger scale tilt is bluer than the smaller scale tilt. Apart from supporting the P18+ACT ($\ell>1800$) analysis findings, P18lowE+ACT locates the upper limit $k_{\rm break}$ till a blue tilt is preferred by ACT. However, if the break in the spectrum occurs at scales larger than $k_{\rm break}>0.158/{\rm Mpc}$, $n_{s2}>n_{s1}$ is preferred at 2$\sigma$ C.L. Therefore, at $k_{\rm break}>0.158/{\rm Mpc}$ or $\ell>2200$, the ACT data does not support a blue spectrum. Here, only using low-$\ell$ EE data from {\it Planck} (that fixes the overall amplitude), we find a 2$\sigma$ preference for $\Delta n_{s}>0$ from ACT. This crucial result indicates that the apparent 3$\sigma$ tension between {\it Planck} PR3 and ACT DR4 is principally driven by the ACT DR4 itself.

    \item {\bf P18-100GHz ($\ell<650$)+ACT}: While P18lowE+ACT analysis finds the small scale limit for the blue tilt preference, it is not possible to find the large scale limit since we did not use any data at large scale to constrain the spectral tilt. P18-100GHz ($\ell<650$)+ACT uses only 100GHz data from {\it Planck} and restricts it to $\ell<650$. Since ACT starts at $\ell=600$ (bin centre) in TT, it minimizes overlap with Planck.~\autoref{Fig:P18100-ell650-ACT} plots the results of this analysis. Here, a clear division exists between the larger scale and the smaller scale tilts. We find $n_{s1} = 0.9616^{+0.0085}_{-0.011}$ and $n_{s2} = 1.000^{+0.039}_{-0.017}$. The scale of change of spectral tilt is also well constrained with $\log_{10}[k_{\rm break}] = -1.11\pm 0.30$. The mismatch between the two tilts in this case identifies the tension between {\it Planck} and ACT as we find $\Delta n_s = 0.038^{+0.048}_{-0.015}$. The mean value of $k_{\rm break}=0.077/{\rm Mpc}$ indicates the mean divisive scale in the power spectrum. The power spectrum (and its residual) posteriors in the bottom panel indicates that at $k\sim0.08/{\rm Mpc}$ (corresponding to $\ell\sim1100$) the joint P18-100GHz ($\ell<650$)+ACT constrained power spectrum starts deviating from P18 mean power law. Combined with the results from P18lowE+ACT we can conclude that the Planck/ACT discrepancy mostly lies within multipoles $1100-2200$. Using Gaussian process, such discrepancy between ACT DR4 and {\it Planck} was found in~\cite{Calderon:2023obf} (see Figure 12 of that publication).

    \item {\bf P18 ($\ell<650$)+ACT}: Here we have used all temperature and polarization auto and cross spectra and restricted the multipoles to $\ell<650$~\footnote{While in~\cite{Smith:2022hwi}, in one of the analysis reported, the {\it Planck} temperature data was truncated to $\ell<650$, in our analysis we truncate both temperature and polarization data to minimize overlap.}. The results are plotted and compared with P18-100GHz ($\ell<650$)+ACT results in~\autoref{Fig:P18100-ell650-ACT}. An increase in the significance for non-zero $\Delta n_s$ also indicates the internal consistency within {\it Planck} spectra at these multipoles and mismatch with ACT at smaller scales. In this analysis the tilts are better constrained with $n_{s1} = 0.9606^{+0.0073}_{-0.0089}$
  and $n_{s2} = 0.994^{+0.029}_{-0.0078}$. The significance of the non-zero difference in tilt increases with $\Delta n_{s}=0.0331^{+0.033}_{-0.0077}$ that indicates the tension between these two datasets. A few samples extending to $\Delta n_{s}<0$ regions are only allowed by the data if the transition occurs at a very large scale or at a very small scale.  
 
     \item {\bf P18-143-217GHz+ACT ($\ell>1800$)}: \autoref{Fig:P18143217-ACT} demonstrates the analysis where {\it Planck} 100GHz channel is not used. As in P18+TT analysis, here too ACT TT data is used after $\ell=1800$ to avoid the small scale overlap. In this case we expect to obtain results similar to P18+ACT ($\ell>1800$). The power spectrum posterior is very similar to P18+ACT ($\ell>1800$) except at very large and small scales. $\Delta n_s - n_{s1}$ plot indicates that $\Delta n_s<0$ is driven by the large scale tilt being blue at scales larger than $\log_{10}[k_{\rm break}]<-2$ (for $k_{\rm break}<0.01$). The absence of 100 GHz spectrum and the power suppression driven by the low multipole temperature anisotropy prefer a blue tilt at largest scales. The power suppression at large scales can also be noticed in the panels at the bottom.
   
    \item {\bf P18EE+ACT}: In this analysis we only use polarization auto-correlation spectrum from Planck. Complete ACT TT data is used here. The results of the analysis, plotted in~\autoref{Fig:P18EE-ACT} can be treated as a consistency check. Certain parameter posteriors are similar to P18 ($\ell<650$)+ACT analysis. This result supports the consistency of {\it Planck} TT and EE data at large scales. Since low-$\ell$ temperature data is not used in this analysis, we find the large scale tilt $n_{s1}<1$ at all scales and $\Delta n_s>0$ is strongly supported at all scales. 
    
    \item {\bf P18TEEE+ACT}: Similar to P18EE+ACT this analysis also acts as a consistency check and support for the change in the tilt in transition from larger to smaller scales. Posteriors in~\autoref{Fig:P18TEEE-ACT} are very similar to the ones plotted in~\autoref{Fig:P18EE-ACT} that again refers the consistency between {\it Planck} temperature and polarization data. 
\end{enumerate} 

\autoref{tab:comparison} contains the differences in best fit $\chi^2_{\rm eff}$ between the broken power law and baseline. As we mentioned in the text discussing the data combinations, here we notice that when {\it Planck} small scale data is not included and ACT data from all bandpowers are used, the broken power law become significant in identifying the tension. We obtain $\Delta\chi^2_{\rm eff}\simeq 7-10$ with 2 extra parameters from the broken power law model. We find consistent trends in the difference between the Bayesian evidences that is reflected in the Bayes' factor tabulated in the third column. 

\begin{table}[!htb]
\begin{center}    
\begin{tabular}{|l|l|l|}
\hline
Datasets                         & $\Delta \chi^2_{\rm eff}$ & $\ln B$ \\ \hline
P18+ACT                          & -2.4                      & -0.3    \\ \hline
P18lowE+ACT                      & -6.8                      & 0.7     \\ \hline
P18-100GHz ($\ell<650$)+ACT      & -8.1                      & 0.3     \\ \hline
P18 ($\ell<650$)+ACT             & -7.1                      & 1.2     \\ \hline
P18-143-217GHz+ACT ($\ell>1800$) & -0.7                       & -1.1    \\ \hline
P18EE+ACT                        & -9.7                      & 1.8     \\ \hline
P18TEEE+ACT                      & -6                        & 0.9     \\ \hline
\end{tabular}
 \caption{The difference in $\chi^2_{\rm eff}$ between the broken power law model and the baseline obtained from analyses with different dataset combinations are tabulated in second column. Third column contains the logarithm of the ratio of Bayesian evidences or the Bayes' factor. Note that a $\chi^2_{\rm eff}$ (also a positive $\ln~B$) indicates a support for the transition in the tilt compared to the baseline.}
    \label{tab:comparison}
\end{center}
\end{table}
\paragraph{Background Parameters}
\begin{figure}[!htb]
\includegraphics[width=0.9\textwidth]{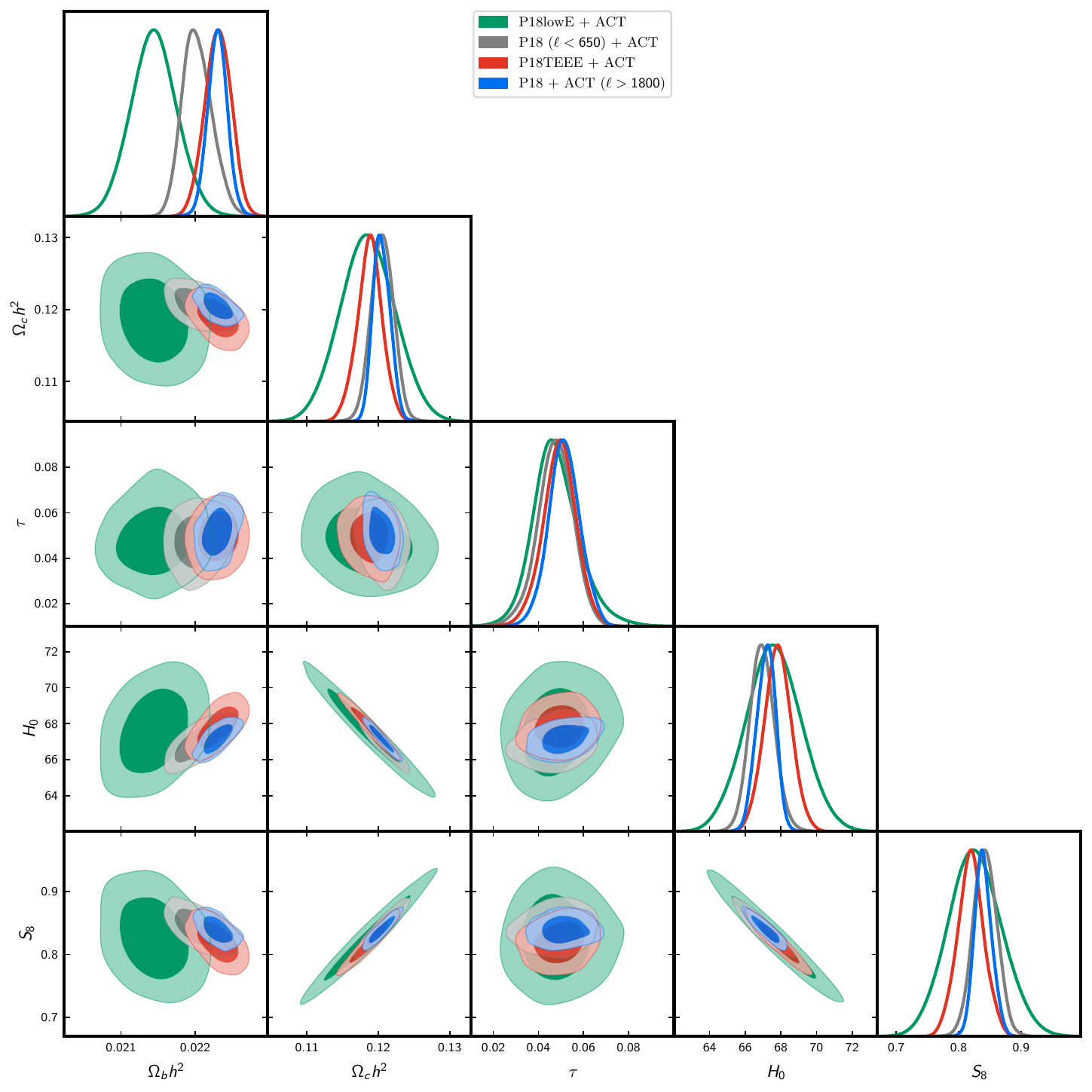}
\caption{Constraints on the background cosmological parameters obtained from different dataset combinations.}~\label{Fig:Background}
\end{figure}

We plot the background parameter constraints in~\autoref{Fig:Background} from some of the crucial data combinations we used. Apart from baryon density, we find consistency in all parameter constraints. The baryon density constraint in ACT DR4 analysis~\cite{ACT:2020} also conflicts with the {\it Planck} and WMAP constraints. The baryon density and the spectral index are inversely correlated and in ACT DR4 analysis a D220 prior (temperature power spectrum prior at $\ell=220$ from WMAP/Planck) or using an overall {\it ad hoc} scaling of TE spectrum was used to move the $\Omega_b h^2 - n_s$ contour to the direction of {\it Planck} constraints. Note that in the first case, use of the prior D220 is expected to degrade the fit to the small scale data from ACT since it forces the spectrum to have a red tilt. Our analysis finds the transition of the spectral index within the ACT data which results in an improvement in fit (as reflected from the $\Delta n_s$ significance) to the data combination. We, at the same time, as expected, find the inverse correlation between $\Omega_b h^2 - n_{s2}$. Therefore we expect, a correction in the ACT data can lead to a resolution to the $\Omega_b h^2 - n_s$ discrepancy and provide a more realistic explanation of the tension compared to the use of large scale angular spectrum priors. The baryon density shifts significantly towards baseline P18 value in the analysis with P18 ($\ell < 650$)+ ACT. This result indicates that also large scale measurement from ACT will be helpful in partially resolving this discrepancy with Planck. 

\subsection{Results from the non-parametric reconstruction}

\begin{figure}[!htb]
\includegraphics[width=0.9\textwidth]{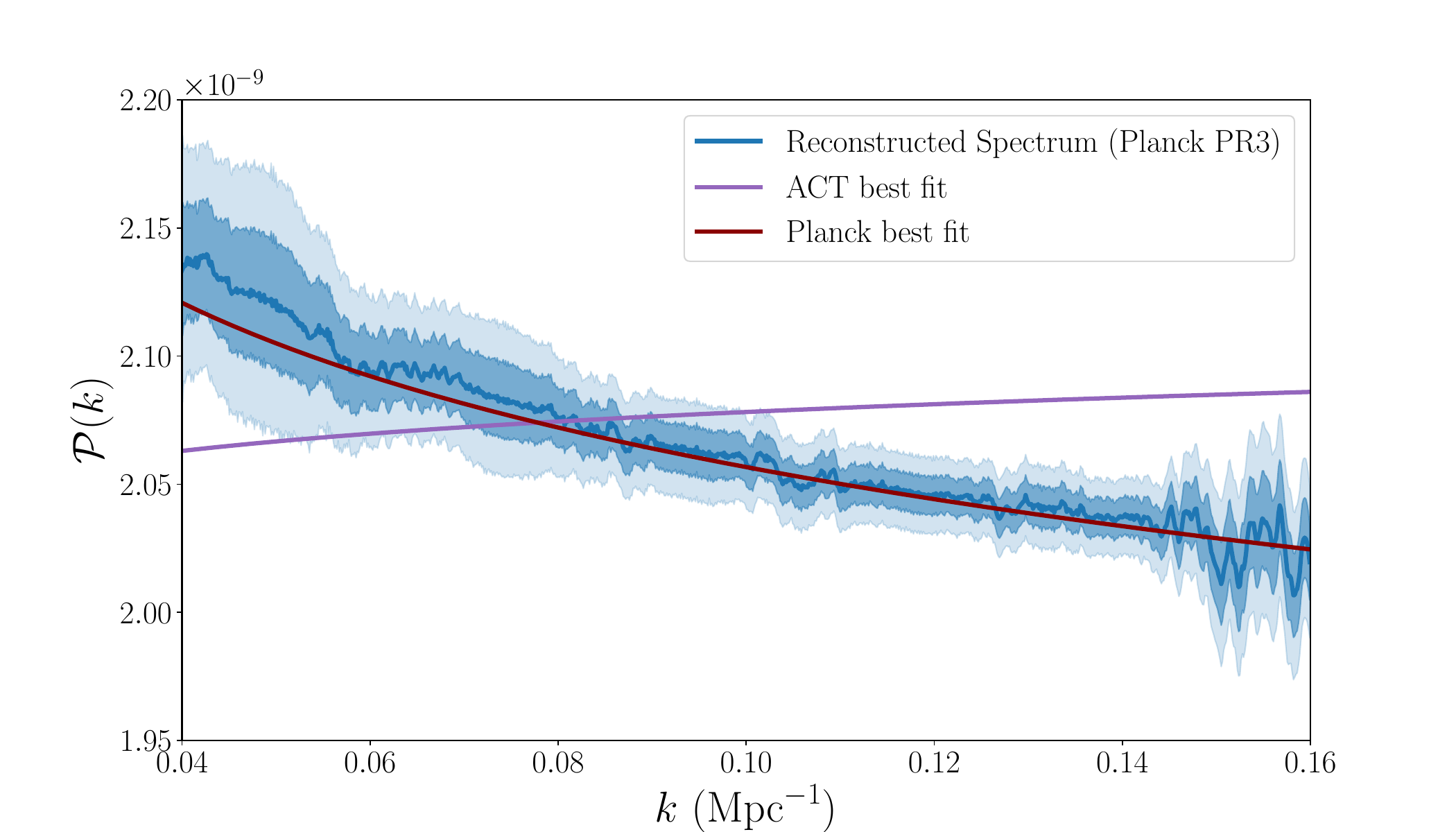}
\includegraphics[width=0.9\textwidth]{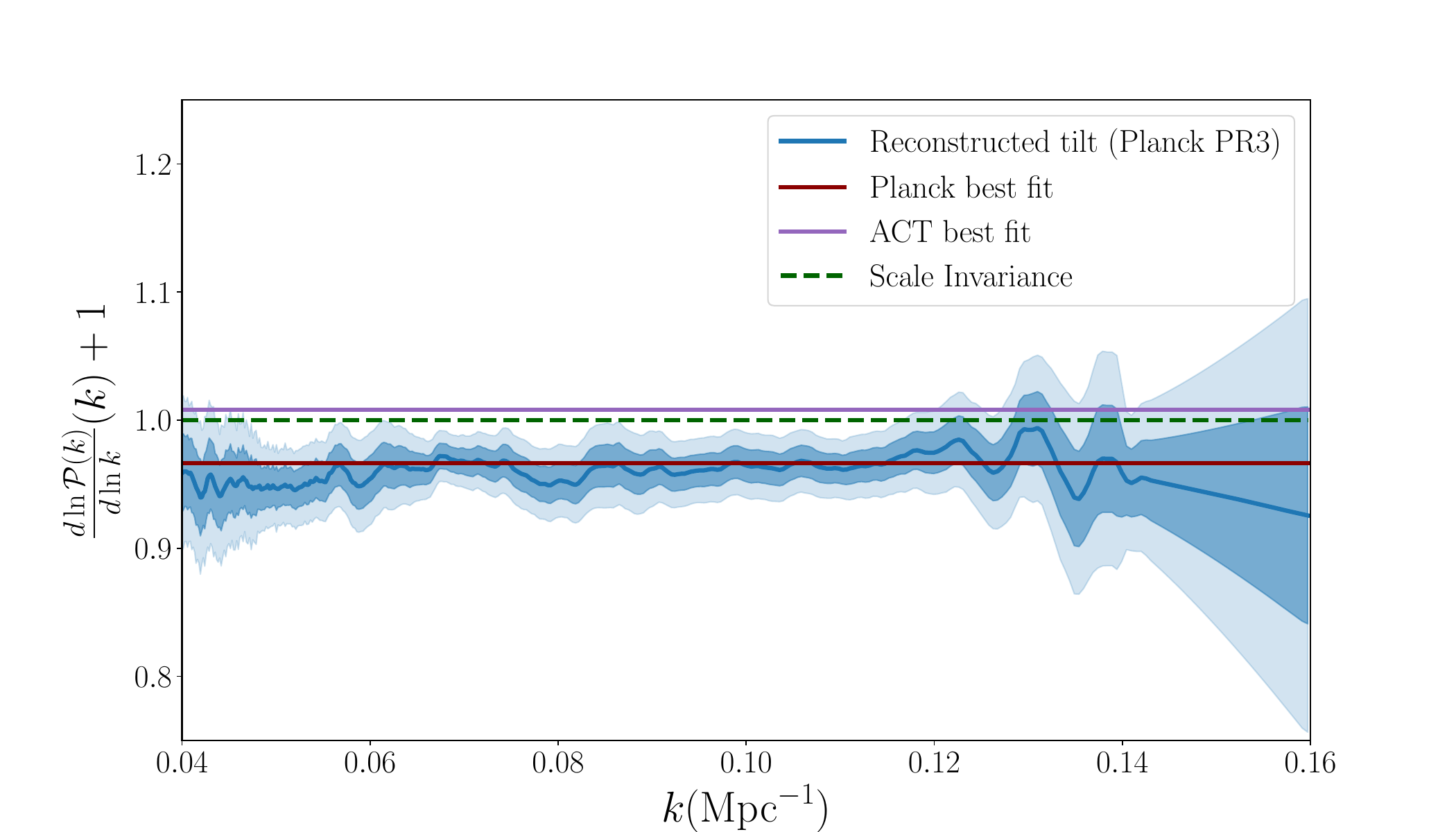}
\caption{Results from primordial power spectrum reconstruction from {\it Planck} PR3. {\it Planck} and ACT best fit references here are plotted for the best fit baseline model. At all scales the baseline model is consistent with the reconstructed power spectrum band (top panel). Reconstructed effective tilts deviate strongly from scale invariant spectrum and remain completely consistent with $n_s\simeq0.96$. At scales smaller than 0.12 ${\rm Mpc}^{-1}$, due to the decrease in the signal-to-noise ratio, the uncertainty on the reconstructed tilt increases.}~\label{Fig:Reconstruction-Planck}
\end{figure}

We plot the results from the primordial spectrum reconstruction from {\it Planck} in~\autoref{Fig:Reconstruction-Planck}. The top panel in the figure contains the 1 and 2$\sigma$ error band around the reconstructed spectrum in darker and lighter shades respectively. The {\it Planck} baseline best fit power spectrum remains within the 1$\sigma$ error band. This consistency indicates remarkable agreement of the power law model with the data. While short-length correlated features with moderate significance can be extracted from the data with higher iterations~\cite{HazraMRLPlanck:2014,Sohn:2022jsm}, in this project, we restrict ourselves to fewer iterations, considering the necessity of comparing the wide features between {\it Planck} and ACT. The ACT baseline best fit is also plotted that shows strong discrepancies compared to the reconstructed band from Planck. The bottom panel plots the effective spectral tilt ($n_s^{\rm eff}$) reconstructed from data. It is numerically computed from each of the reconstructed spectrum from all realizations of the data. The best fit baseline tilts from {\it Planck} PR3 and ACT DR4 also also provided with the scale invariant reference. The $n_s^{\rm eff}$ strongly rejects a scale invariant spectra between scales 0.05-0.12 ${\rm Mpc}^{-1}$. At this point it is again important to highlight that the features around $\ell=750-850$ as discussed in~\cite{HazraMRLPlanck:2014} require significant changes in the spectral tilt at scales localized around 0.06 ${\rm Mpc}^{-1}$. This is not evident here as with fewer iterations the fine features are not reconstructed in this work, and only certain changes, hinted around the aforementioned scale, can be observed. Note that, the reconstructions from scales smaller than 0.04~${\rm Mpc}^{-1}$ are shown here to compare with ACT reconstruction in the next analysis.

\begin{figure}[!htb]
\includegraphics[width=0.9\textwidth]{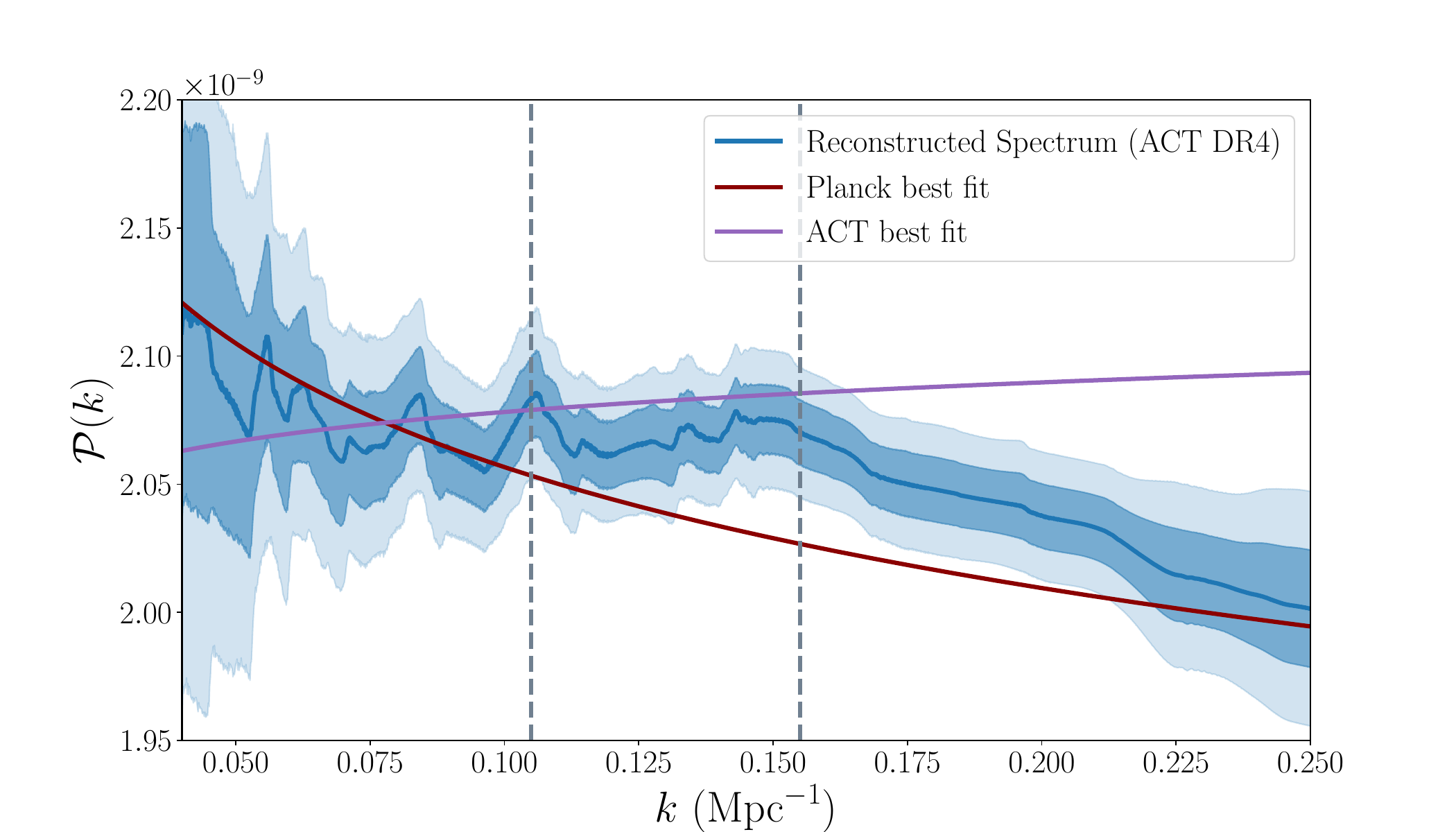}
\includegraphics[width=0.9\textwidth]{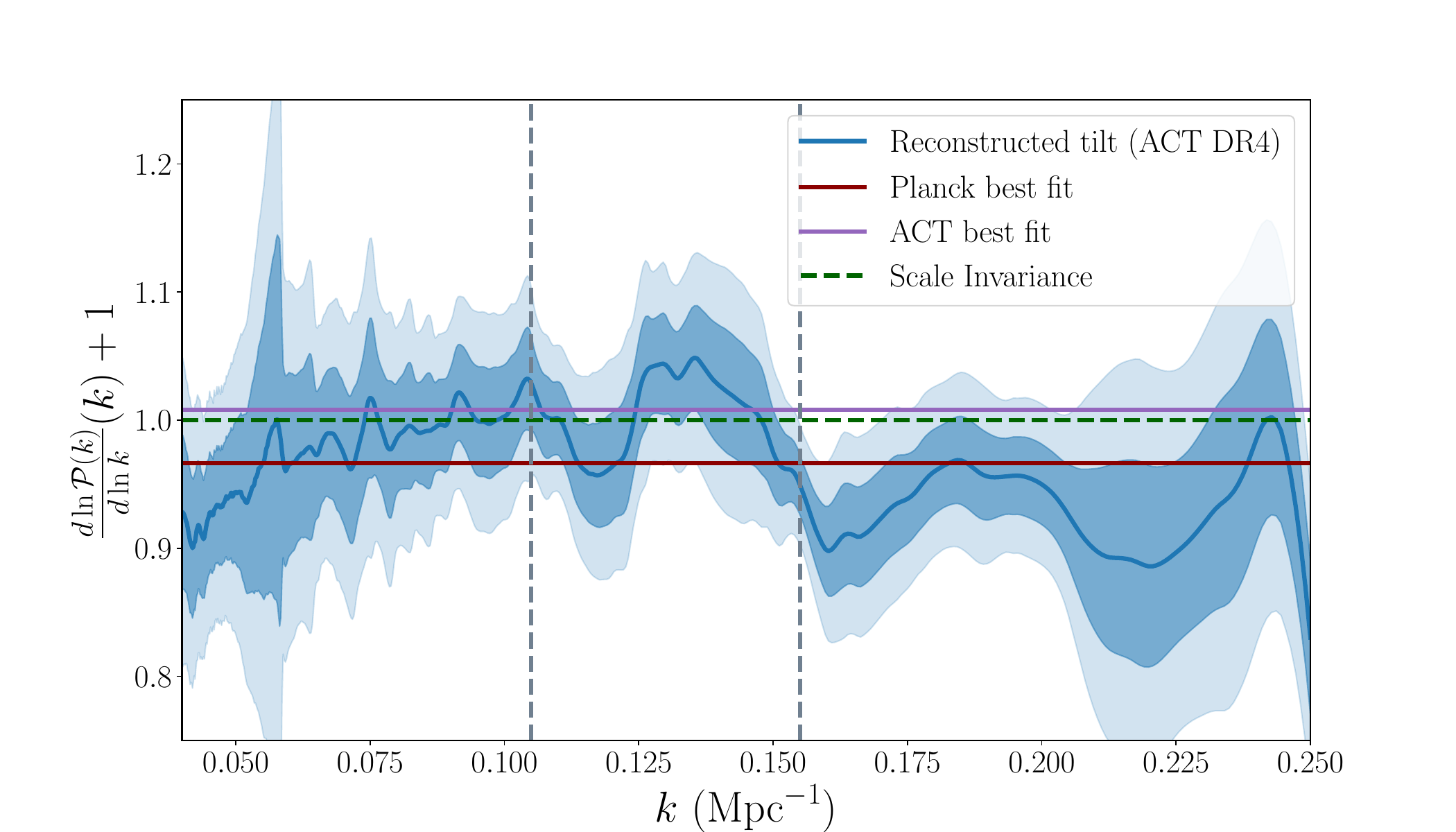}
\caption{Results from primordial power spectrum reconstruction from {\it Planck} PR3 + ACT DR4. As discussed in~\autoref{subsec:method-reconstruction}, the reconstruction uses co-added unbinned {\it Planck} data till multipole 574 and ACT data afterwards. The plot at the top show the difference between {\it Planck} best fit and the reconstructed band (1 and 2$\sigma$ represented by darker and lighter bands respectively). The effective tilt is plotted in the bottom panel with the best fit spectral indices from {\it Planck} and ACT and the scale invariance. The vertical line at 0.105${\rm Mpc}^{-1}$ indicate the beginning of deviation of the reconstructed band from the {\it Planck} baseline best fit tilt (comparing with the plot at the bottom panel) at more than 2$\sigma$ significance. The vertical line at 0.155${\rm Mpc}^{-1}$ marks the scale where the reconstructed tilt becomes red and at 2$\sigma$ significance rejects the scale invariance. } \label{Fig:Reconstruction-ACT}
\end{figure}

The results of primordial spectrum reconstruction using {\it Planck} ($\ell\le574$) and ACT ($\ell\ge575$, the central multipole for the bin being $\ell=600$) data are plotted in~\autoref{Fig:Reconstruction-ACT}. The top and bottom panel here correspond to the plots of reconstructed power spectrum and effective tilt bands. Before 0.04~${\rm Mpc}^{-1}$, the reconstruction is dominated by {\it Planck} data and therefore we do not plot this here. In this free form reconstruction, the difference with {\it Planck} baseline best fit can be clearly identified. Till 0.1~${\rm Mpc}^{-1}$ the spectrum contains oscillations with both blue and red tilts at different scales. However, the {\it Planck} baseline best fit remains consistent at 2$\sigma$
at all scales with the reconstructed band. After 0.1~${\rm Mpc}^{-1}$ the spectrum becomes nearly scale invariant with a blue tilt. Between 0.105 and 0.155 ${\rm Mpc}^{-1}$ reconstructed band significantly differs from the {\it Planck} best fit baseline tilt (notice the plot at the bottom panel of the figure). At scales smaller than 0.155 ${\rm Mpc}^{-1}$, the spectrum turns red and at more than 2$\sigma$ it rejects a scale invariant spectrum. These results are consistent with what we have obtained in the broken power law model analysis. We would like to note that, while at 0.155 ${\rm Mpc}^{-1}$ the spectrum becomes red, the power spectrum plot still show differences {\it w.r.t.} {\it Planck} best fit line because of the enhanced amplitude of the PPS. Around 0.2 ${\rm Mpc}^{-1}$ the reconstructed spectrum becomes consistent with {\it Planck} best fit again.

\subsection{Possibility of new physics vs. systematics}
 Exploring the detailed impact of instrumental and astrophysical systematics on the constraints presented in the previous sections is beyond the scope of this paper. However, in the next paragraphs we list a couple of such nuisances. 

\paragraph{Foreground mismodelling} This work makes use of the ``foreground marginalised'' ACT likelihood. Errors in the modelling of foreground power spectra and frequency dependence could lead to excess or lack of power on small scales. Besides, ACT and {\it Planck} plik likelihoods use slightly different foreground models. Yet, we expect these effects to be very small compared to the sensitivity of both observations.

\paragraph{Scale dependent instrumental systematics} Instrumental systematics could also artificially alter the slope of the power spectra, leading to the observed discrepancies between the two data sets. In particular, an error in modelling either the beam or the sky model during the mapmaking step~\cite{Naess:2023} could lead to potentially large instrumental systematics that are not trivial to marginalise over.

A quantitative estimation of the impact of such instrumental and astrophysical systematics is left to a future paper. 

\subsection{Sky variation}
\begin{figure}
    \centering
    \includegraphics[width=\textwidth]{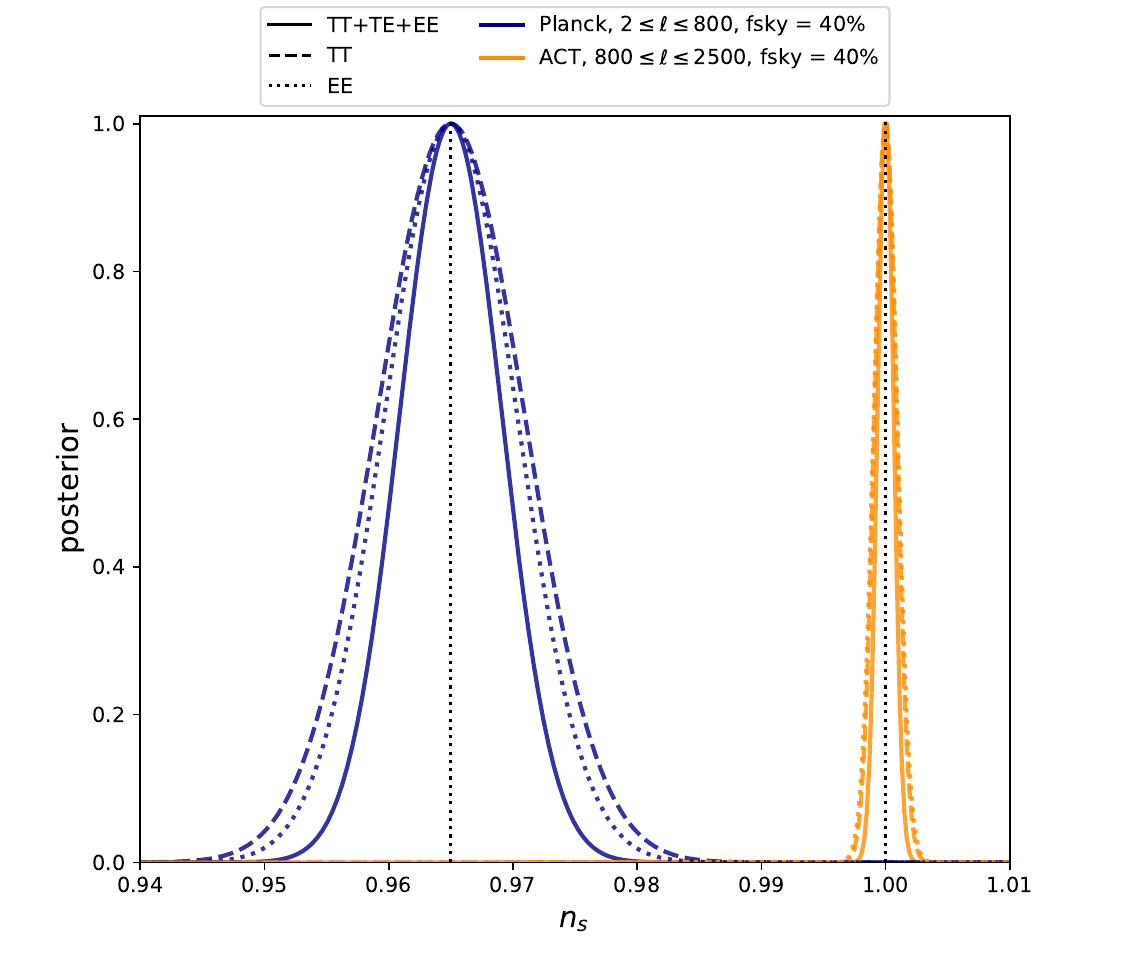}
    \caption{Posteriors associated to a cosmic variance-limited measurement of $n_s=0.965$ (resp. $1.0$) on a $f_{\rm sky}=40\%$, mimicking a noiseless {\it Planck} (resp. ACT) observation. }
    \label{fig:Planck_vs_ACT_cosmic_variance}
\end{figure}

In~\autoref{fig:Planck_vs_ACT_cosmic_variance} we provide an estimate of the possibility of having different spectral tilts at different scales and at different sky patches. Here, assuming cosmic variance limited observation, we provide the posterior distribution on spectral tilt expected from the 40\% of the sky (patch not covered by ACT DR4) from the power spectra between multipoles $2<\ell\le800$, and from the rest of the sky (the other 40\% corresponding to the patch covered by ACT DR4) between multipoles $800<\ell\le2500$. For the former, we assume a fiducial $n_s=0.965$ and for the other we assume $n_s=1$. We find that for a cosmic variance limited observation, having two tilts at two different patches will create a highly significant statistical discrepancy. In fact, the result highlight that, for a comic variance limited temperature anisotopy observation (such as Planck) such differences in mean $n_s$ in two patches would result in significant internal tension. We do not however, explore this {\it Planck} consistency using ACT mask since in our analysis we identified that the departure from a red tilt in ACT spectrum is localized to certain scales only and within the ACT data the difference in tilt is supported at a moderate level (2$\sigma$).

\section{Summary}~\label{sec:summary}
We explore the tension between {\it Planck} PR3 and ACT DR4 in the determination of the spectral tilt using two approaches -- parametric modelling and non-parametric reconstruction. In the parametric modelling we have used a broken-power law model that allows two spectral tilts in the primordial spectrum around a variable cosmological length scale. In the non-parametric reconstruction, we use Modified-Richardson-Lucy deconvolution algorithm to reconstruct the primordial power spectrum. These two approaches are complementary as the parametric approach finds broad features/differences in the data marginalized over the background cosmological parameters while the deconvolution brings out the fine features in the data for a fixed background cosmology. We have used different combinations of {\it Planck} and ACT data to understand the constraints on the spectral tilt from data at different cosmological scales. Below, we provide the key results obtained in our analysis.
\begin{enumerate}
    \item In the parametric analysis, we find a moderate significance (about 2$\sigma$) for the break in the primordial spectrum (transition in the spectral tilt) from ACT DR4 only data. The use of {\it Planck} EE, TEEE and large-to-intermediate scale ($\ell=2-650$) TTTEEE data increases the significance of the transition of the spectral tilt from red (larger scales) to blue (smaller scales).
    \item Parametric reconstruction also highlights that at scales smaller than $k\sim0.16/{\rm Mpc}$ ($\ell>2200$), the ACT data prefers a red tilt. Joint analysis with {\it Planck} data truncated at $\ell<650$ indicates that the preference for the blue tilt by ACT starts at $k> 0.08/{\rm Mpc}$ ($\ell>1100$). These analyses highlight that ACT spectrum is not in agreement with the {\it Planck} spectrum after $k\sim0.08/{\rm Mpc}$ ($\ell\sim1100$) but again strongly supports a red tilt after $k\sim0.16/{\rm Mpc}$ ($\ell\sim2200$) at more than 2$\sigma$ significance.  
    
    \item The Bayes' factors obtained between the baseline model and the broken power law model highlight that the broken power law model is favored only when {\it Planck} temperature data is not used or truncated at scales $\ell<650$. The improvement from the broken power law model degrades when ACT data is used from $\ell>1800$.   

    \item The non-parametric reconstruction explores the finer features beyond the spectral tilt. Our reconstruction highlights a blue-ward change in the spectral tilt in the ACT data between $k\sim0.11-0.16/{\rm Mpc}$ which is consistent with our finding from the parametric analysis. The reconstruction from {\it Planck} remains completely consistent with the best fit power law spectrum (with a red-tilt) at all scales.

    \item The differences in the baryon density estimation between {\it Planck} PR3 and ACT DR4 seems to be
    \begin{enumerate}
        \item partially due to the less sky coverage of ACT that masks the first peak of spectrum. A combined analysis with {\it Planck} PR3 ($\ell<650$) + ACT DR4 where we avoid overlap in the datasets,  shifts the baryon density significantly higher towards the {\it Planck} only analysis,
        
        \item contributed by the intermediate transition of spectral tilt.        
    \end{enumerate}

    therefore, we believe both large scale measurement and small scale correction would be needed in resolving this baryon density discrepancy.

\end{enumerate}

Given the large overlap between {\it Planck} PR3 and ACT DR4 in the cosmological scales where we have found the deviation ($k\sim0.08-0.16/{\rm Mpc}$) from a single power law spectrum, it appears that it is difficult to accommodate the spectral tilt tension between {\it Planck} and ACT with new physics that is compatible with both datasets. This is also supported by our findings where the preference for the new model (here, explored with the broken power law) increases only when the {\it Planck} temperature data is not used at small scales. The unavailability of the first acoustic peak in the ACT DR4 data is not responsible for the spectral tilt tension as only with ACT data we find the preference for a change in tilt at 95\% C.L. Therefore, foreground mismodelling or scale dependent instrumental systematics can play major roles in this tension. With ACT DR6 data release, we expect to understand the CMB data and model consistency with much more detail.

\section*{Acknowledgements}
The authors acknowledge the use of computational resources
at the Institute of Mathematical Science’s High Performance Computing facility [Kamet and Nandadevi]. DKH would like to acknowledge the support from CEFIPRA grant no. 6704-4 and India-Italy mobility program (INT/Italy/P-39/2022 (ER)). BB and JE acknowledge the SCIPOL project funded by the European Research Council (ERC) under the European Union’s Horizon 2020 research and innovation program (PI: Josquin Errard, Grant agreement No. 101044073). AS would like to acknowledge the support by National Research Foundation of Korea NRF2021M3F7A1082056 and the support of the Korea Institute for Advanced Study (KIAS) grant funded by the government of Korea. The authors would like to thank Fabio Finelli, Daniela Paoletti and Adrien La Posta
for their comments on this manuscript.

\bibliography{PlanckACT}

\providecommand{\noopsort}[1]{}\providecommand{\singleletter}[1]{#1}%

\providecommand{\href}[2]{#2}\begingroup\raggedright\begin{thebibliography}{10}

\bibitem{ACT:2020}
{\scshape ACT} collaboration, \emph{{The Atacama Cosmology Telescope: DR4 Maps
  and Cosmological Parameters}},
  \href{https://doi.org/10.1088/1475-7516/2020/12/047}{\emph{JCAP} {\bfseries
  12} (2020) 047} [\href{https://arxiv.org/abs/2007.07288}{{\ttfamily
  2007.07288}}].

\bibitem{Planck:2018param}
{\scshape Planck} collaboration, \emph{{Planck 2018 results. VI. Cosmological
  parameters}},
  \href{https://doi.org/10.1051/0004-6361/201833910}{\emph{Astron. Astrophys.}
  {\bfseries 641} (2020) A6}
  [\href{https://arxiv.org/abs/1807.06209}{{\ttfamily 1807.06209}}].

\bibitem{Handley:2020hdp}
W.~Handley and P.~Lemos, \emph{{Quantifying the global parameter tensions
  between ACT, SPT and Planck}},
  \href{https://doi.org/10.1103/PhysRevD.103.063529}{\emph{Phys. Rev. D}
  {\bfseries 103} (2021) 063529}
  [\href{https://arxiv.org/abs/2007.08496}{{\ttfamily 2007.08496}}].

\bibitem{DiValentino:2022rdg}
E.~Di~Valentino, W.~Giar\`e, A.~Melchiorri and J.~Silk, \emph{{Quantifying the
  global \textquoteleft{}CMB tension\textquoteright{} between the Atacama
  Cosmology Telescope and the Planck satellite in extended models of
  cosmology}}, \href{https://doi.org/10.1093/mnras/stad152}{\emph{Mon. Not.
  Roy. Astron. Soc.} {\bfseries 520} (2023) 210}
  [\href{https://arxiv.org/abs/2209.14054}{{\ttfamily 2209.14054}}].

\bibitem{Planck:2016tof}
{\scshape Planck} collaboration, \emph{{Planck intermediate results. LI.
  Features in the cosmic microwave background temperature power spectrum and
  shifts in cosmological parameters}},
  \href{https://doi.org/10.1051/0004-6361/201629504}{\emph{Astron. Astrophys.}
  {\bfseries 607} (2017) A95}
  [\href{https://arxiv.org/abs/1608.02487}{{\ttfamily 1608.02487}}].

\bibitem{Hazra:2013broken}
D.~K. Hazra, A.~Shafieloo and G.~F. Smoot, \emph{{Reconstruction of broad
  features in the primordial spectrum and inflaton potential from Planck}},
  \href{https://doi.org/10.1088/1475-7516/2013/12/035}{\emph{JCAP} {\bfseries
  12} (2013) 035} [\href{https://arxiv.org/abs/1310.3038}{{\ttfamily
  1310.3038}}].

\bibitem{Richardson:72}
W.~H. Richardson, \emph{Bayesian-based iterative method of image
  restoration$\ast$}, \href{https://doi.org/10.1364/JOSA.62.000055}{\emph{J.
  Opt. Soc. Am.} {\bfseries 62} (1972) 55}.

\bibitem{Lucy:74}
L.~B. {Lucy}, \emph{{An iterative technique for the rectification of observed
  distributions}}, \href{https://doi.org/10.1086/111605}{\emph{Astronomical
  Journal} {\bfseries 79} (1974) 745}.

\bibitem{HazraMRLWMAP:2013}
D.~K. Hazra, A.~Shafieloo and T.~Souradeep, \emph{{Primordial power spectrum: a
  complete analysis with the WMAP nine-year data}},
  \href{https://doi.org/10.1088/1475-7516/2013/07/031}{\emph{JCAP} {\bfseries
  07} (2013) 031} [\href{https://arxiv.org/abs/1303.4143}{{\ttfamily
  1303.4143}}].

\bibitem{HazraMRLPlanck:2014}
D.~K. Hazra, A.~Shafieloo and T.~Souradeep, \emph{{Primordial power spectrum
  from Planck}},
  \href{https://doi.org/10.1088/1475-7516/2014/11/011}{\emph{JCAP} {\bfseries
  11} (2014) 011} [\href{https://arxiv.org/abs/1406.4827}{{\ttfamily
  1406.4827}}].

\bibitem{Shafieloo_2012}
A.~Shafieloo, \emph{Crossing statistic: Bayesian interpretation, model
  selection and resolving dark energy parametrization problem},
  \href{https://doi.org/10.1088/1475-7516/2012/05/024}{\emph{Journal of
  Cosmology and Astroparticle Physics} {\bfseries 2012} (2012) 024–024}.

\bibitem{Hazra:2013oqa}
D.~K. Hazra and A.~Shafieloo, \emph{{Test of consistency between Planck and
  WMAP}}, \href{https://doi.org/10.1103/PhysRevD.89.043004}{\emph{Phys. Rev. D}
  {\bfseries 89} (2014) 043004}
  [\href{https://arxiv.org/abs/1308.2911}{{\ttfamily 1308.2911}}].

\bibitem{Hazra:2014hma}
D.~K. Hazra and A.~Shafieloo, \emph{{Confronting the concordance model of
  cosmology with Planck data}},
  \href{https://doi.org/10.1088/1475-7516/2014/01/043}{\emph{JCAP} {\bfseries
  01} (2014) 043} [\href{https://arxiv.org/abs/1401.0595}{{\ttfamily
  1401.0595}}].

\bibitem{Shafieloo:2016zga}
A.~Shafieloo and D.~K. Hazra, \emph{{Consistency of the Planck CMB data and
  $\Lambda$CDM cosmology}},
  \href{https://doi.org/10.1088/1475-7516/2017/04/012}{\emph{JCAP} {\bfseries
  04} (2017) 012} [\href{https://arxiv.org/abs/1610.07402}{{\ttfamily
  1610.07402}}].

\bibitem{Hazra:2014rulingout}
D.~K. Hazra, A.~Shafieloo, G.~F. Smoot and A.~A. Starobinsky, \emph{{Ruling out
  the power-law form of the scalar primordial spectrum}},
  \href{https://doi.org/10.1088/1475-7516/2014/06/061}{\emph{JCAP} {\bfseries
  06} (2014) 061} [\href{https://arxiv.org/abs/1403.7786}{{\ttfamily
  1403.7786}}].

\bibitem{Handley:CC}
W.~J. Handley, M.~P. Hobson and A.~N. Lasenby, \emph{Polychord: next-generation
  nested sampling}, \href{https://doi.org/10.1093/mnras/stv1911}{\emph{Monthly
  Notices of the Royal Astronomical Society} {\bfseries 453} (2015)
  4385–4399}.

\bibitem{gitCosmoChord}
W.~J. Handley, ``{G}it{H}ub - williamjameshandley/{C}osmo{C}hord:
  {C}osmological sampling with {P}oly{C}hord + {C}osmo{M}{C} --- github.com.''
  \url{https://github.com/williamjameshandley/CosmoChord}.

\bibitem{esaPlanckLegacy}
``{P}lanck {L}egacy {A}rchive --- pla.esac.esa.int.''
  \url{https://pla.esac.esa.int}.

\bibitem{actpollite}
``{L}{A}{M}{B}{D}{A} - {A}{C}{T} ({A}{C}{T}{P}ol) {D}ata {P}roducts - {D}{R}4
  {C}osmological {R}esults {D}ownloads --- lambda.gsfc.nasa.gov.''
  \url{https://lambda.gsfc.nasa.gov/product/act/act_dr4_cosmic_results_get.html}.

\bibitem{ShafielooIRL:2004}
A.~Shafieloo and T.~Souradeep, \emph{{Primordial power spectrum from WMAP}},
  \href{https://doi.org/10.1103/PhysRevD.70.043523}{\emph{Phys. Rev. D}
  {\bfseries 70} (2004) 043523}
  [\href{https://arxiv.org/abs/astro-ph/0312174}{{\ttfamily
  astro-ph/0312174}}].

\bibitem{PLA}
``{P}lanck {L}egacy {A}rchive --- pla.esac.esa.int.''
  \url{http://pla.esac.esa.int/pla/#home}.

\bibitem{Calderon:2023obf}
R.~Calder\'on, A.~Shafieloo, D.~K. Hazra and W.~Sohn, \emph{{On the consistency
  of \ensuremath{\Lambda}CDM with CMB measurements in light of the latest
  Planck, ACT and SPT data}},
  \href{https://doi.org/10.1088/1475-7516/2023/08/059}{\emph{JCAP} {\bfseries
  08} (2023) 059} [\href{https://arxiv.org/abs/2302.14300}{{\ttfamily
  2302.14300}}].

\bibitem{Smith:2022hwi}
T.~L. Smith, M.~Lucca, V.~Poulin, G.~F. Abellan, L.~Balkenhol, K.~Benabed
  et~al., \emph{{Hints of early dark energy in Planck, SPT, and ACT data: New
  physics or systematics?}},
  \href{https://doi.org/10.1103/PhysRevD.106.043526}{\emph{Phys. Rev. D}
  {\bfseries 106} (2022) 043526}
  [\href{https://arxiv.org/abs/2202.09379}{{\ttfamily 2202.09379}}].

\bibitem{Sohn:2022jsm}
W.~Sohn, A.~Shafieloo and D.~K. Hazra, \emph{{Free-form reconstruction of
  primordial power spectrum using Planck CMB temperature and polarization}},
  \href{https://arxiv.org/abs/2211.15139}{{\ttfamily 2211.15139}}.

\bibitem{Naess:2023}
S.~Naess and T.~Louis, \emph{Large-scale power loss in ground-based cmb
  mapmaking}, \href{https://doi.org/10.21105/astro.2210.02243}{\emph{The Open
  Journal of Astrophysics} {\bfseries 6} (2023) }.

\end{thebibliography}\endgroup

\end{document}